\documentclass{aa}
\usepackage[english]{babel}

\usepackage{amsmath}
\usepackage{physics}
\usepackage{cleveref}[2012/02/15]
\crefformat{footnote}{#2\footnotemark[#1]#3}

\usepackage{xcolor}
\usepackage[normalem]{ulem} 
\usepackage{graphicx}
\usepackage{txfonts}
\usepackage{xurl}
\newcommand{\ind}[1]{_{\rm #1}}
\newcommand{\super}[1]{^{\rm #1}}
\begin{document}

   \title{Plasma densities, flow and Solar EUV flux at comet 67P}
   \subtitle{A cross-calibration approach}

   \author{F. L. Johansson
          \inst{1,2},
           A. I. Eriksson \inst{1}
        \and
        E. Vigren \inst{1}
        \and
        L Bucciantini\inst{3}
        \and
           P. Henri \inst{3,4}
        \and
             H. Nilsson \inst{5}
        \and
            \\ S. Bergman \inst{5,6}
        \and
            N. J. T. Edberg \inst{1}
        \and
        G. Stenberg Wieser \inst{5}
        \and
        E. Odelstad \inst{1}
          }

   \institute{Swedish Institute of Space Physics,
              Uppsala, Sweden\\
              \email{frejon@irfu.se}
         \and
             Uppsala University, Department of Astronomy and Space Physics, Uppsala, Sweden
             \and Laboratoire de Physique et Chimie de l'Environnement et de l'Espace, CNRS, Orl\'eans, France
            \and Laboratoire Lagrange, OCA, CNRS, UCA, Nice, France
            \and Swedish Institute of Space Physics, Kiruna, Sweden
            \and Ume\aa ~University, Department of Physics, Ume\aa, Sweden
             }

   \date{Received XX; accepted YY}
   \authorrunning{F. L. Johansson et al.}

\abstract
{During its two-year mission at comet 67P, Rosetta nearly continuously monitored the inner coma plasma environment for gas production rates varying over three orders of magnitude, at distances to the nucleus from a few to a few hundred km. To achieve the best possible measurements, cross-calibration of the plasma instruments is needed.}
{The goal is to provide a consistent plasma density data set for the full mission, in the process providing a statistical characterisation of the plasma in the inner coma and its evolution.}
   {We construct physical models for two different methods to cross-calibrate the spacecraft potential and the ion current as measured by the Rosetta Langmuir Probes (LAP) to the electron density as measured by the Mutual Impedance Probe (MIP). We also describe the methods used to estimate spacecraft potential, and validate the results with the Ion Composition Analyser, (ICA).}
   {We retrieve a continuous plasma density dataset for the entire cometary mission with a much improved dynamical range compared to any plasma instrument alone and, at times, improve the temporal resolution from 0.24-0.74~Hz to 57.8~Hz. The physical model also yields, at 3~hour time resolution, ion flow speeds as well as a proxy for the solar EUV flux from the photoemission from the Langmuir Probes.}
   {We report on two independent mission-wide estimates of the ion flow speed which are consistent with the bulk H$_2$O$^+$ ion velocities as measured by ICA. We find the ion flow to consistently be much faster than the neutral gas over the entire mission, lending further evidence that the ions are collisionally decoupled from the neutrals in the coma. RPC measurements of ion speeds are therefore not consistent with the assumptions made in previously published plasma density models of the comet ionosphere at the start and end of the mission. Also, the measured EUV flux is perfectly consistent with independently derived values previously published from LAP and lends support for the conclusions drawn regarding an attenuation of solar EUV from a distant nanograin dust population, when the comet activity was high. The new density dataset is consistent with the existing MIP density dataset, but facilitates plasma analysis at much shorter timescales, and covers also long time periods where densities were too low to be measured by MIP.}

   \keywords{plasmas -- comets:individual: 67P/Churyumov-Gerasimenko -- space vehicles: instruments, methods: data analysis, methods:statistical}

   \maketitle
%

\section{Introduction}

ESA's comet chaser Rosetta studied the comet 67P/Churyumov-Gerasimenko in unprecedented detail for more than two years from August 2014 to September 2016 \citep{taylor_rosetta_2017}. In what amounted to a half-revolution around the sun, the scientific package dedicated to the plasma environment, the \emph{Rosetta Plasma Consortium} \citep[RPC,][]{carr_rpc:_2007}, studied the comet evolve through different activity levels including perihelion at 1.24~AU and the dwindling activity at the end of the mission, at 3.83~AU. The RPC included, among other instruments, the \emph{Langmuir probe} \citep[LAP,][]{eriksson_rpc-lap:_2007,eriksson_cold_2017}, the \emph{Ion Composition Analyzer} \citep[ICA,][]{nilsson_rpc-ica:_2007,nilsson_evolution_2017} and the \emph{Mutual Impedance Probe} \citep[MIP,][]{trotignon_rpc-mip:_2007,henri_diamagnetic_2017}.

\begin{figure}
\centering
    \includegraphics[width=0.9\columnwidth]{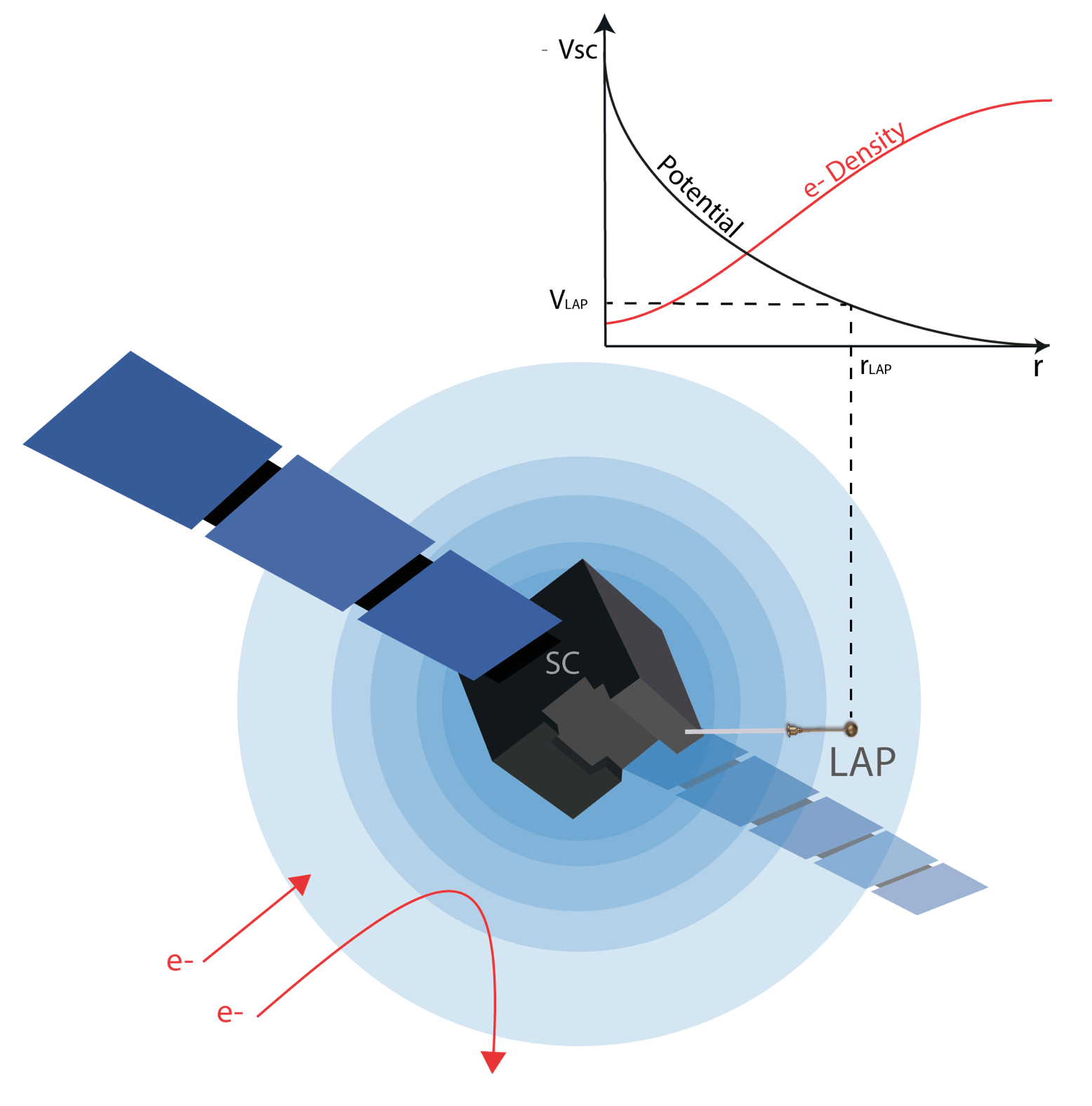}
    \caption{Sketch showing the effect on electron density at the Langmuir Probe (LAP1) position from the electrostatic field from the (negative) spacecraft potential that envelopes the sensor. Equipotentials of the electrostatic potential field in the same plane as LAP1 and the spacecraft centre are visualised by concentric blue circles.}
    \label{fig:rosetta_sketch}
\end{figure}

As the spacecraft became significantly negatively charged in the cometary environment \citep{odelstad_evolution_2015,odelstad_measurements_2017, johansson_charging_2020}, the charged particles that constitute the plasma environment were attracted to or repelled from the spacecraft and instruments mounted on the spacecraft body or on short booms protruding from the spacecraft body, as sketched in Figure~\ref{fig:rosetta_sketch}. Low energy particles, i.e.\ with energy in eV comparable to the spacecraft potential in volts, are particularly affected: cometary electrons are significantly repelled from the Langmuir Probe \citep{eriksson_cold_2017,johansson_charging_2020}, and positive ions are perturbed \citep{bergman_influence_2019,bergman_influence_debye_2020}. As such, this poses significant challenges for RPC's ability to characterise the low energy cometary plasma which dominates the comet environment \citep{edberg_spatial_2015,odelstad_ion_2018,gilet_mutual_2019,wattieaux_plasma_2020}, but can be overcome through accurate modelling \citep{bergman_influence_2019,johansson_charging_2020} and cross-calibration \citep{heritier_vertical_2017,Breuillard2019xcal} to MIP measurements, which have been found to be almost unperturbed by a negative spacecraft potential \citep{wattieaux_rpcmip_model_2019}.

We structure this paper as follows: After a brief description of the instruments used in this study in Section~\ref{sec:instrument}, we describe the spacecraft potential as measured by LAP, and how we can verify that measure using the attracted ions observed by ICA in Section~\ref{sec:scpot}. We then show how the spacecraft potential obstruct some of the LAP measurement modes, but also how we can use the spacecraft potential to our advantage to recover electron densities via cross-calibration in Section~\ref{sec:NEL}, and in a similar fashion cross-calibrate the LAP ion currents to electron densities. We thereby obtain a dataset that combines the dynamic range and high resolution of LAP with the accuracy of MIP. Finally, we present a side-effect of the physical model of the cross-calibration routine, which yields accurate estimates on the ion flow speed and EUV flux and agree with independent measures and previously published results.

\section{Instrument description}
\label{sec:instrument}

The Rosetta plasma consortium included, among other instruments, the \emph{Rosetta dual Langmuir probe instrument} (LAP), which consists of two 5~cm diameter spherical electrostatic probes LAP1 and LAP2 and their associated electronics. The primary parameter measured are the currents flowing to (or the voltage of) the probes when some bias voltages (or bias currents) are applied to it. From these measurements we can derive plasma characteristics such as plasma density, electron temperature, ion flow speed, as well as spacecraft potential, extreme ultraviolet (EUV) flux and electric field fluctuations. Not all parameters are monitored simultaneously or uniformly over the mission, as it is contingent on the bias voltage or current (i.e. the operational mode) selected for each probe, and, ultimately, the science objective of each operational period.
 LAP1 and LAP2 are situated at the ends of separate booms of length 2.24 and 1.62~m, respectively, and described in further detail in \citet{eriksson_rpc-lap:_2007}.

The Mutual Impedance Probe, (MIP), is an active electric sensor that allows us to derive the plasma density via the identification of the plasma resonance frequency. MIP consists of two pairs of transmitting and receiving electrodes (as well as their associated electronics), and measures the electric coupling of the electrodes to the plasma by fluctuating a charge on the transmitters. MIP is mounted on a 1~m long bar on the same boom as LAP1, but can also make use of the LAP2 sensor for transmission, situated $\sim 4$~m away on the LAP2 boom. MIP is described in further detail by \citet{trotignon_rpc-mip:_2007}.

The Ion Composition Analyzer, ICA, is a mass-resolving ion spectrometer mounted on the Rosetta spacecraft body, and included in the RPC package. ICA has a maximum energy range of a few eV to 40~keV, which it typically sweeps every 12~s  with a typical energy resolution of $dE/E = 0.07$, as well as an angular field-of-view of 360 $\times$ 90~$\deg$~\citep{odelstad_measurements_2017,nilsson_evolution_2017}. As positive ions are accelerated into the ICA instrument when the spacecraft potential is negative, we can estimate the spacecraft potential also from the measured energy of the lowest energy ions, as described in \citet{odelstad_measurements_2017}.

\begin{figure}
    \includegraphics[width=1.0\columnwidth]{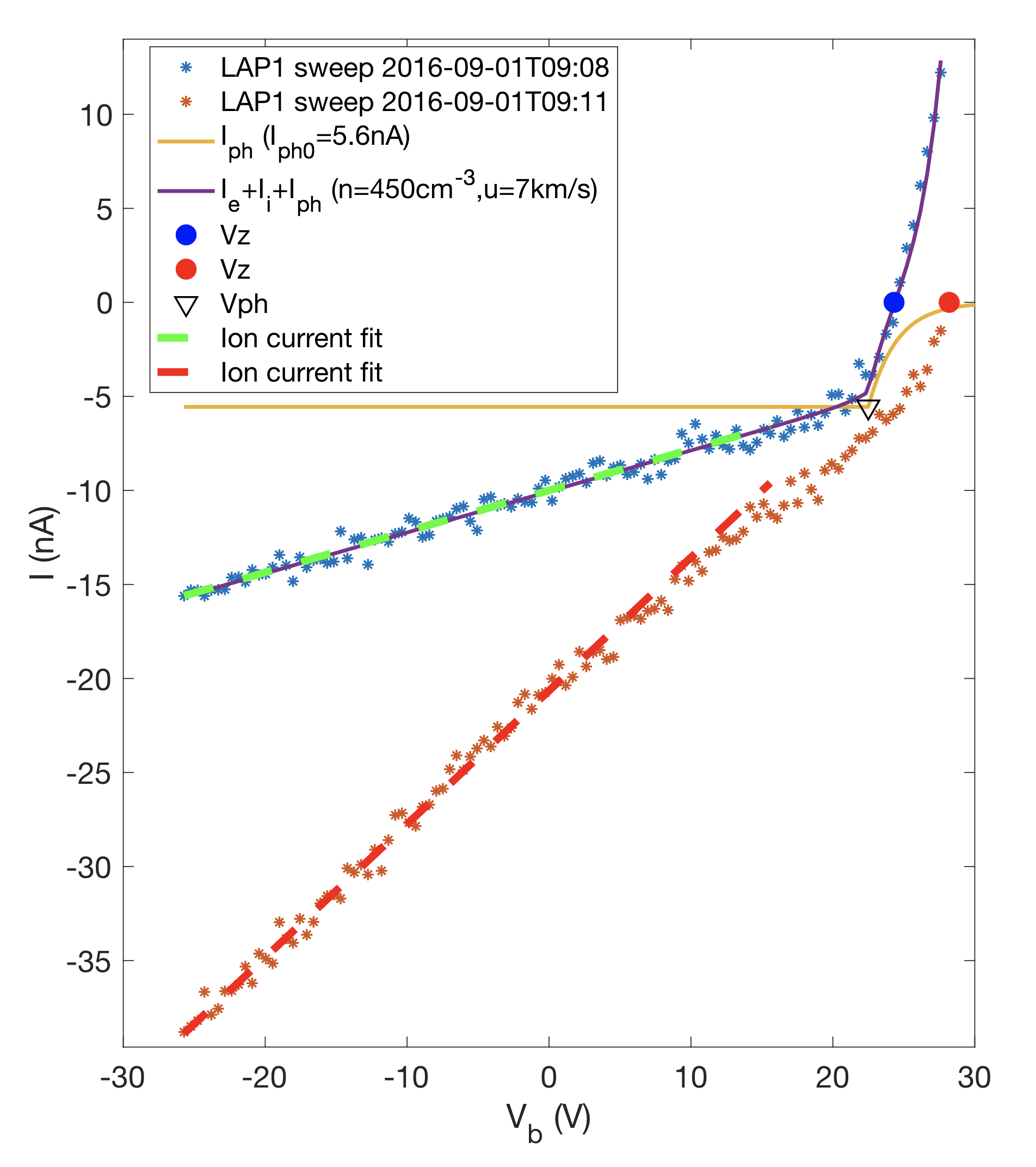}
    \caption{Two example LAP sweeps from 2016-09-01 (blue and red stars) showing significant spacecraft charging, as well as a visualisation of a selection of parameters from the analysis: the spacecraft potential estimate from the zero-crossing potential, $V\ind{z}$ from each sweep (blue and red dots, respectively), the estimated photoemission current for the first sweep (yellow line), which has a discontinuity at $V\ind{ph}$ (black triangle), the ion current slope fits (green and red dashed lines) and the total current fit to the first sweep (purple solid line). As the spacecraft charging is significant, the electron current is poorly constrained by the bias range, but is here modelled using a double Maxwellian distribution with characteristic energies of 0.1~eV and 1.5~eV.}
    \label{fig:sweepexample}
\end{figure}

\section{Spacecraft potential}\label{sec:scpot}

The spacecraft potential, $V\ind{S}$, is fundamental to the interpretation of plasma data. As documented in detail by \citet{odelstad_evolution_2015,odelstad_measurements_2017}, LAP uses two complementary techniques of estimating the spacecraft potential. One involves identifying the photoelectron knee potential $V\ind{ph}$ from a probe sweep, which is the negative of the potential where the probe is in equilibrium with its immediate plasma surroundings and is neither attracting nor repelling photoelectrons emitted from the probe surface. The other method involves measuring the floating potential $V\ind{f}$, which is the negative of the potential for which the sum of all currents to the probe is zero. When LAP is operating in floating potential mode, we measure $V\ind{f}$ directly, but we can also estimate it from a sweep (Figure~\ref{fig:sweepexample}), in which case we denote this parameter $V\ind{z}$ .

In an ideal scenario, $V\ind{z}$ and $V\ind{f}$ would be fully equivalent. However, the sweep has discrete step size, 0.25 or 0.5~V in most modes, and $V\ind{z}$ has to be found by interpolation or fitting. Therefore, $V\ind{z}$ usually exhibits more noise than does $V\ind{f}$, which is also immune to a displacement current added by any capacitive effects when varying the bias voltage. Nevertheless, the method of fitting $V\ind{z}$ can increase the range of the estimate by extrapolating beyond the sweep bias voltage window, as shown in Figure~\ref{fig:sweepexample}. In one of the two sweeps in this example, $V\ind{z}$ falls just outside the bias range of the sweep, but as we show below, the extrapolation can reach much further. For sweeps with disturbances (noise or otherwise) there can be several zero-crossings of current. In these cases, each zero-crossing is ranked in descending order of longevity, i.e.\ the distance to the next zero-crossing in either direction, as we expect disturbances to be short-lasting over sweep time scales. Only the best ranked zero-crossing is chosen for the $V\ind{z}$ estimate.

\begin{figure}
    \includegraphics[width=1.0\columnwidth]{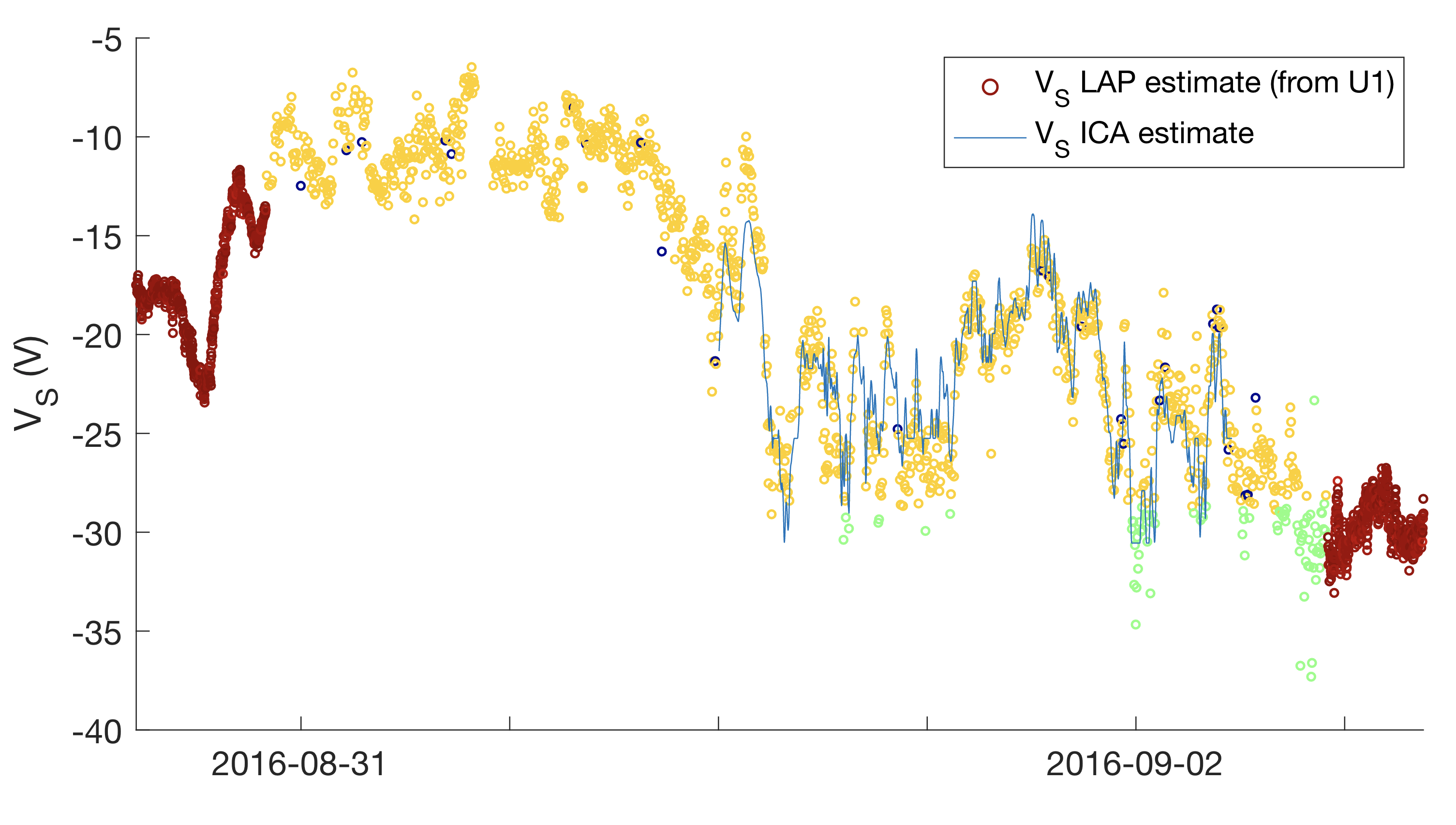}
    \caption{Spacecraft potential estimates from LAP and ICA. Coloured circles are the spacecraft potential estimates from LAP, coloured by data source. Red circles are derived from $V\ind{f}$. Yellow, blue and green circles are all derived from $V\ind{z}$, but green indicates values that are extrapolated beyond the sweep range and blue indicates sweeps with several zero-crossings (i.e.\ disturbances). The blue line is an -3.8~V offset corrected ICA estimate derived from the negative of the lowest energy bin with at least five ion detections, and filtered using a 50 point moving average. Note that there are no obvious discontinuities in moving from $V\ind{z}$ to $V\ind{f}$.}
    \label{fig:vzvf_ica}
\end{figure}

In Figure~\ref{fig:vzvf_ica} we show a validation of the automatic identification of $V\ind{z}$, beyond the sweep range, and in comparison to $V\ind{f}$ and ICA spacecraft potential estimates. No obvious discontinuities are seen between the two methods in Figure~\ref{fig:vzvf_ica}. Statistically, as LAP alternated between continuous floating potential mode and sweeps (yielding $V\ind{f}$ and $V\ind{z}$ estimates, respectively) over the entire mission, we find a slight shift with a median of 0.4~V and a median absolute deviation of 1.7~V, when we move from one mode to the other.

\begin{figure}
    \includegraphics[width=1.0\columnwidth]{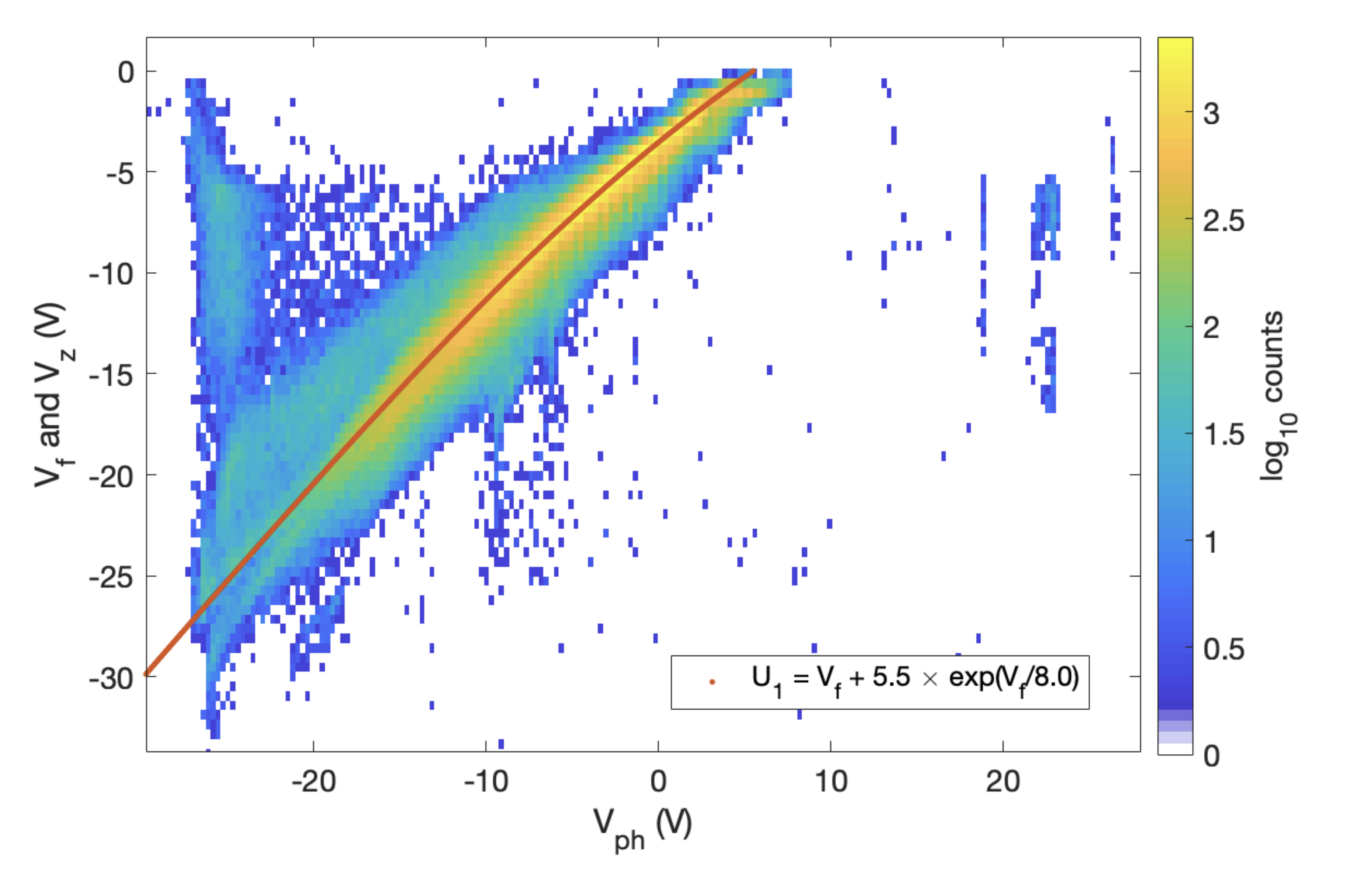}
    \caption{2D histogram of 390~000 simultaneous spacecraft potential estimates $V\ind{ph}$ vs $V\ind{z}$ or $V\ind{f}$ in 200x200 bins, corresponding to bin widths of~$\sim$~0.3 and 0.5~V, coloured by log$_{10}$ counts for the entire cometary mission. As $V\ind{ph}$ is sensitive to noise, there are several artefacts at e.g.\ +22~V and -25~V, but there is otherwise a clear agreement, especially for large negative values as far as $V\ind{ph}$ can measure. As the spacecraft potential approaches zero, $V\ind{f}$ and $V\ind{z}$ diverges non-linearly from $V\ind{ph}$, the latter being considered the better estimate for the spacecraft potential at these ranges. An empirical model was fitted to move from $V\ind{f}$ and $V\ind{z}$ to $V\ind{ph}$ and is plotted in red.}
    \label{fig:vph_vz_u1model}
\end{figure}

How $V\ind{ph}$ and $V\ind{f}$ relate to the spacecraft potential has been studied in detail in \citet{odelstad_measurements_2017} whenever we can compute simultaneous and good quality estimates from the offset of low energy ions in ICA for a spacecraft that is sufficiently negatively charged. Over the entire mission, $V\ind{ph}$ and $V\ind{f}$ diverge slightly as the spacecraft potential approaches zero or becomes slightly positive, with $V\ind{f}$ becoming non-linearly less sensitive to spacecraft potential changes, as shown in Figure~\ref{fig:vph_vz_u1model}. Also, the method of detection for $V\ind{ph}$ relies on identifying the peak of the second derivative, i.e.\ the discontinuity of the photoelectron current and, as such, relies on a very good signal-to-noise ratio for an accurate detection. When the signal-to-noise ratio is poor, a Blackman filter \citep{magnus_digital_2008} has been applied on the signal before identifying the peak of the second derivative, which can introduce errors and artefacts. However, as $V\ind{ph}$ relates to the spacecraft potential only by some factor~\citep{odelstad_measurements_2017},
we created an empirical model to map the less noise-sensitive $V\ind{f}$ and $V\ind{z}$ dataset to equivalent $V\ind{ph}$ values by use of the fit in Figure~\ref{fig:vph_vz_u1model}. We call this new variable $U_1$ according to
\begin{equation}
    U_1 = V\ind{f} + 5.5\ \exp \left(\frac{V\ind{f}}{8.0} \right),
\end{equation}
where we note that the numerical factors have units of volts and that $U_1$ should be some factor $\alpha$ of the spacecraft potential, $V\ind{S}$, between 0.7 and 1 according to \citet{odelstad_measurements_2017}. For the LAP $V\ind{S}$ estimates plotted in Figure~\ref{fig:vzvf_ica}, this factor $\alpha$ is taken to be 0.95. With this empirical model, 68 percent of all $U_1$ estimates fall within 1~V of $V\ind{ph}$ in Figure~\ref{fig:vph_vz_u1model}.

\section{Method} \label{sec:NEL}

\subsection{MIP-LAP cross-calibration}

As described in Section~\ref{sec:instrument}, the RPC included two instruments targeting the bulk plasma properties, LAP and MIP. The Langmuir probe technique used by LAP is very flexible, enabling access to a broad number of plasma parameters like electron density, electron temperature and ion temperature or flow speed, and also provided integrated solar EUV flux, spacecraft potential and electric field measurements. Other advantages are the broad dynamic range in plasma density (see Section~\ref{sec:discussion}) and the possibility of very high time resolution (down to 53~$\mu$s for LAP). However, perturbations from the spacecraft on the plasma at the probe can sometimes be very large, and disentangling variations of density and temperature is not always possible. The mutual impedance technique used by MIP is entirely different, resting on observing the transmission properties of the plasma for artificially injected oscillating electric fields in the kHz to MHz range when in active mode, or observing natural plasma oscillations when in passive mode. Compared to LAP, dynamic range, maximum time resolution and resolution in density are all lower, but MIP also has two major advantages. First, the main resonance frequency identified by MIP effectively only depends on the plasma density, with very little complication from temperature variations. Second, for the negative spacecraft potential attained by Rosetta in the inner coma, the MIP density value turns out to be insensitive to the spacecraft potential \citep{wattieaux_rpcmip_model_2019}.

In order to combine the advantages of the two instruments and measurement principles as discussed above, cross-calibrated datasets have been made available on the ESA Planetary Systems Archive\footnote{\url{http://psa.esa.int/}}\citep{PSA_2017}. In the following sections we will show some results of these cross-calibrations as well as some by-products achieved during the process. In principle, the cross-calibration uses the current or voltage measured by LAP and finds a fit of this to available MIP density data points over some time period. This fit can then be used either to obtain a plasma density with the high resolution and/or dynamic range of LAP measurements and the robustness of MIP measurements.

\begin{figure}
    \includegraphics[width=0.998\columnwidth]{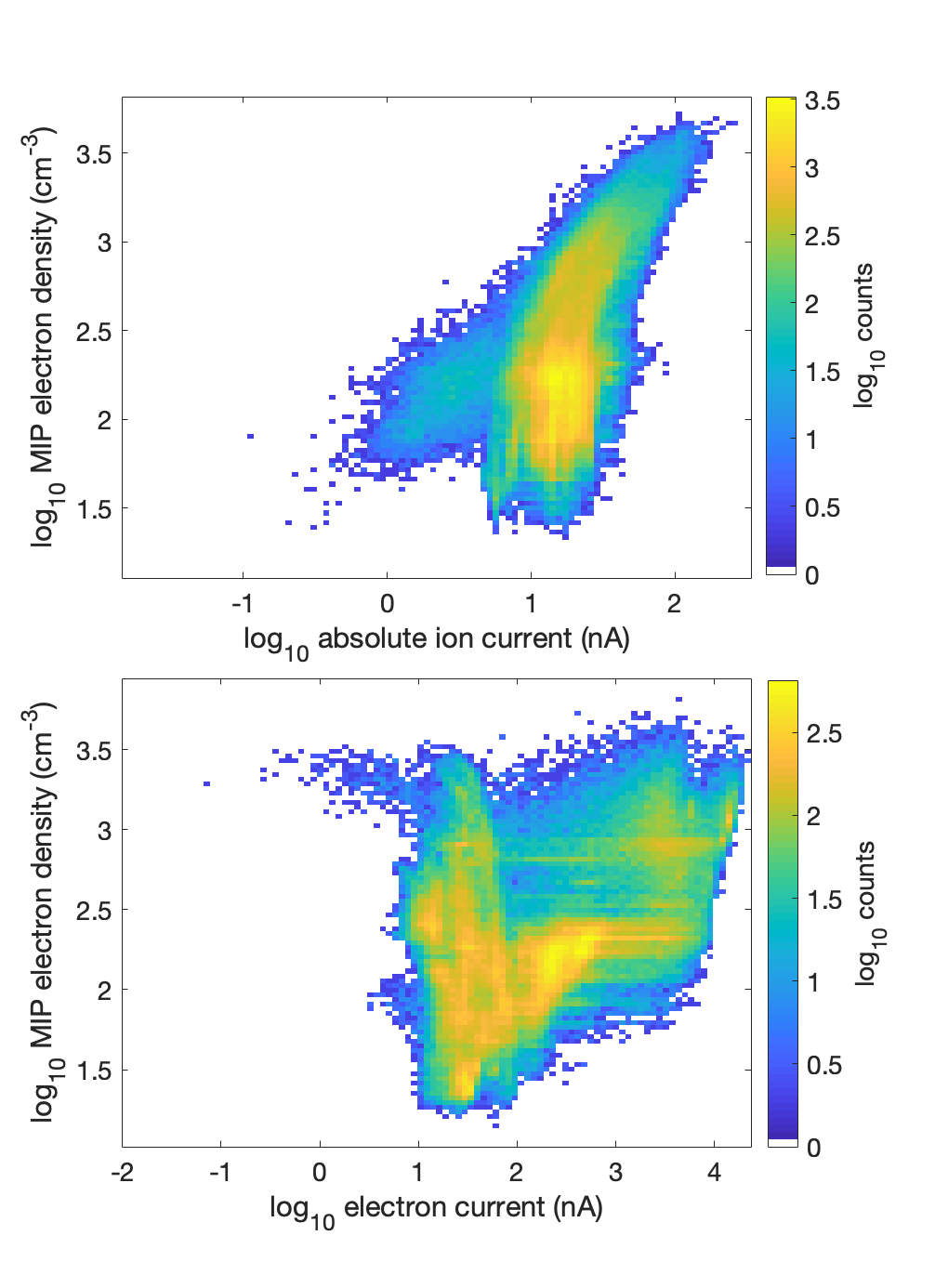}
    \caption{LAP currents in 32-second averages vs simultaneous 32-second average MIP densities over the mission. \textbf{Top:} MIP density vs LAP1 current at negative bias voltage ($V\ind{b}<-15$~V, referred to as "ion current"), coloured by log10 counts in each bin. \textbf{Bottom:} MIP density vs LAP1 electron currents ($V\ind{b}$ >20~V), coloured by log10 counts in each bin.}
    \label{fig:Scandinavia}
\end{figure}

The upper panel in Figure~\ref{fig:Scandinavia} shows a mission-wide 2D histogram of the (32~s average) MIP electron density vs.\ the (32~s average) current measured by one of the two LAP probes, LAP1, at fixed negative bias voltage. For a non-illuminated probe, this is mainly the current due to ion collection, proportional to the plasma density as seen by the points extending down to about 1~nA and extending to the top right corner. However, when the probe is exposed to sunlight the resulting current cannot be lower than the contribution from photoelectron emission, as seen by the large number of points around to 10 to 50~nA. Section~\ref{sec:NEL_I} will show how this contribution can be identified and used for solar EUV monitoring. As described in \citep{johansson_rosetta_2017}, LAP2 has exhibited clear contamination signatures from a capacitive and resistive layer, and as such, is rarely used for current cross-calibration. Nevertheless, when in floating potential mode, there is no electrical response from the contamination layer, and we have excellent agreement between the two probes.

The lower panel of Figure~\ref{fig:Scandinavia} illustrates why only currents at negative bias voltage (ion currents) are used for cross-calibration. This 2D histogram shows currents obtained at positive bias voltage with respect to the spacecraft, intended to attract electrons providing a current proportional to plasma density. While some points indeed show the expected correlation, others clearly do not, even exhibiting anti-correlation, and the spread of points is very large. The basic reason for this is the highly negative potential obtained by the spacecraft in high density plasmas \citep{johansson_charging_2020}, so high that the resulting voltage of the probe can even become negative with respect to the plasma. Such charging is not a problem for the ion current, which maintains a good correlation to the MIP plasma density. This also lends observational support to the insensitivity of the MIP electron density to the spacecraft potential as found by \citet{wattieaux_rpcmip_model_2019}.

We may here note that the spacecraft potential, MIP electron density estimates and LAP currents are all measures of plasma parameters on slightly different spatial scales. The currents to the 5~cm diameter Langmuir probe mainly depends on the plasma within about a meter\footnote{The electrostatic field from the Langmuir probe does not reach much farther than the Debye length, but in practice its reach is limited to about a meter even when the Debye length is longer as the distance at which the vacuum field from a sphere at potential $\sim$30~V has decayed to the typical energy of the plasma particles collected ($\sim$5~eV) is on the order of 10 probe radii.} from the probe and therefore is the most local estimate, while the spacecraft potential (derived from currents to a largely conductive spacecraft with a wingspan of 32~m) is the least local. When there are plasma variations on scales below tens of meters, even ideal sets of measurements would only agree in an average sense. As the Debye length in the inner coma typically is $\lesssim 1$~m \citep{Gilet2019cold}, such small scale plasma variations are fully possible, meaning that the scatter observed when comparing e.g.\ LAP probe current to MIP plasma density is not necessarily due to measurement uncertainties.

\begin{figure}
    \includegraphics[width=0.998\columnwidth]{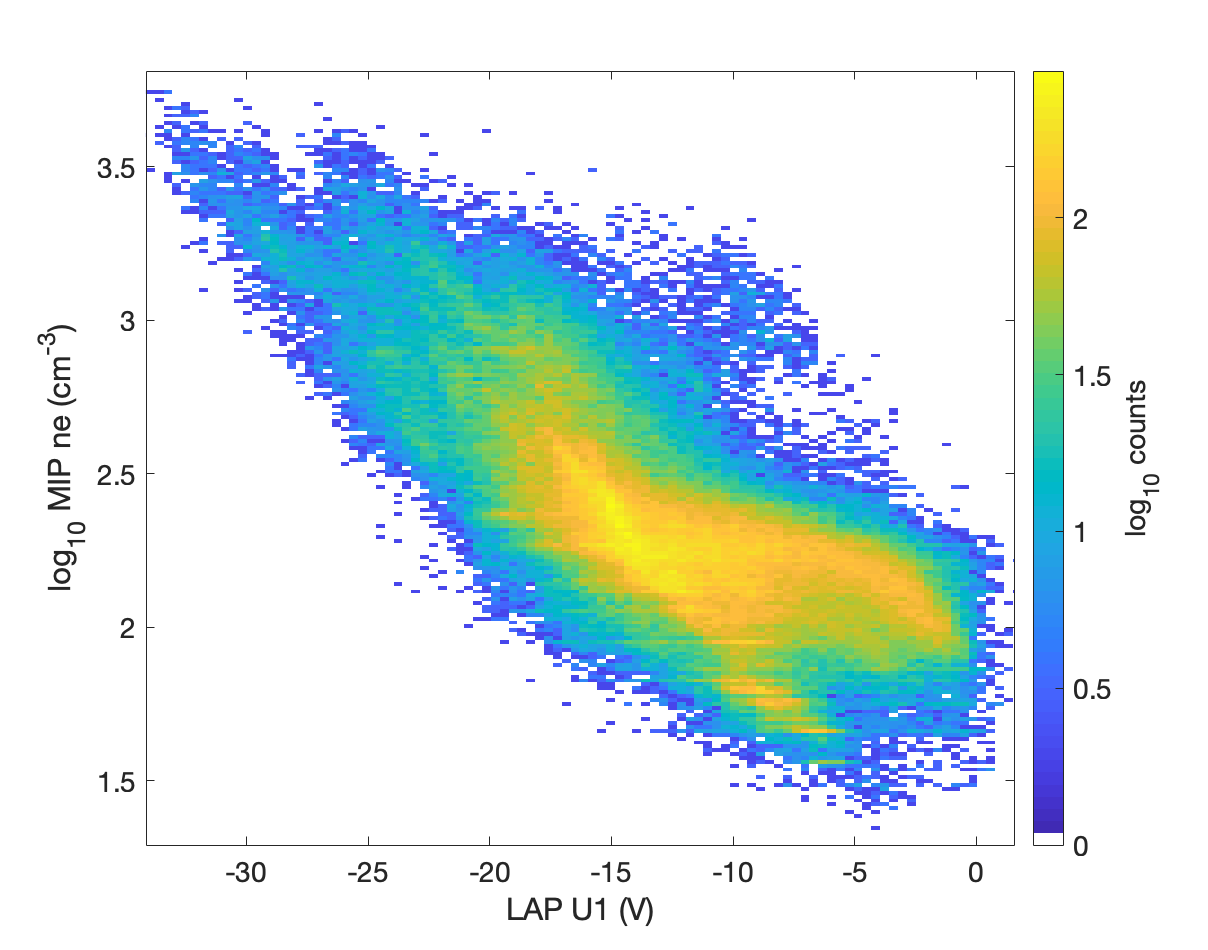}
    \caption{Mission-wide 2-D histogram of MIP density vs simultaneous LAP spacecraft potential proxy $U_1$ in 100x200 bins coloured by log$_{10}$ counts. The upper sensitivity limit of MIP in LDL mode is just below 300~cm$^{-3}$, and is visible as a partial cut-off.}
    \label{fig:mipne_lapu1}
\end{figure}

\subsection{Cross-calibration of spacecraft potential to electron density}
\label{sec:NEL_V}

MIP identifies the electron density via detection of the plasma frequency. Typically, every 1.4-4.3 seconds,
an oscillating electric field is injected in the plasma at stepped frequencies through different electric transmitters. The electric potential that propagates in the plasma is simultaneously measured on the two MIP receivers, from which the complex (amplitude and phase) mutual impedance spectra is built. Eventually, the resonance observed in the mutual impedance spectra at the plasma frequency enables us to retrieve the electron density at a cadence of 0.24-0.74~Hz.

There are two fundamental operational modes of MIP, namely the SDL and LDL modes, used intermittently for smaller and larger Debye lengths, respectively. Therefore, the density targeted by the MIP instrument is limited to two different ranges, and as such, on the selection of a suitable operational mode based on the predicted density days/weeks in advance. It is therefore not surprising that there are intervals in the cometary mission where the electron density falls outside of MIP sensitivity range (see Section~\ref{sec:results}, Figure~\ref{fig:xcal_nednelexamples}). Thus, special care has to be taken when the average density is outside the measurement range but the dynamic variations of the plasma allow MIP to sporadically detect the lower or upper edge of the density range. If a MIP sample (i.e.\ a positive MIP plasma resonance identification) is not adjacent to multiple positive identifications (at least five within the ten closest identifications), the MIP sample is disregarded for cross-calibration.

The spacecraft potential is the potential for which the sum of all currents to the spacecraft is zero. Defining a positive current as the flow of positive charges from the spacecraft, the dominating positive and negative current contributions for Rosetta are the cometary electron current and the photoemission current, respectively. In such a plasma, for a negative and conductive spacecraft, $V\ind{S}$ becomes \citep{odelstad_measurements_2017}
\begin{equation}\label{eq:Vs}
    V\ind{S} \approx -T\ind{e} \log \left( C\, n \sqrt{T\ind{e}}\  \right),
\end{equation}
where $n$ is the number density of the electrons, which are assumed to be a Maxwellian population of characteristic temperature $T\ind{e}$, given in eV, and $C$ is a constant not depending on plasma properties.

We therefore expected the spacecraft potential to be much more sensitive to the temperature of these electrons than to the electron density. In a recent study \citep{johansson_charging_2020}, we have shown that this is not the case, and also presented a detailed model explaining why the Rosetta spacecraft potential is sensitive to the density of electrons, regardless of temperature, allowing even cold (0.1~eV) electrons to reach a spacecraft that is often charged -10 to -20~V. The primary reason is the presence of exposed positively biased conductors on the front-side edges of the solar panels.

\begin{figure}
    \includegraphics[width=1.0\columnwidth]{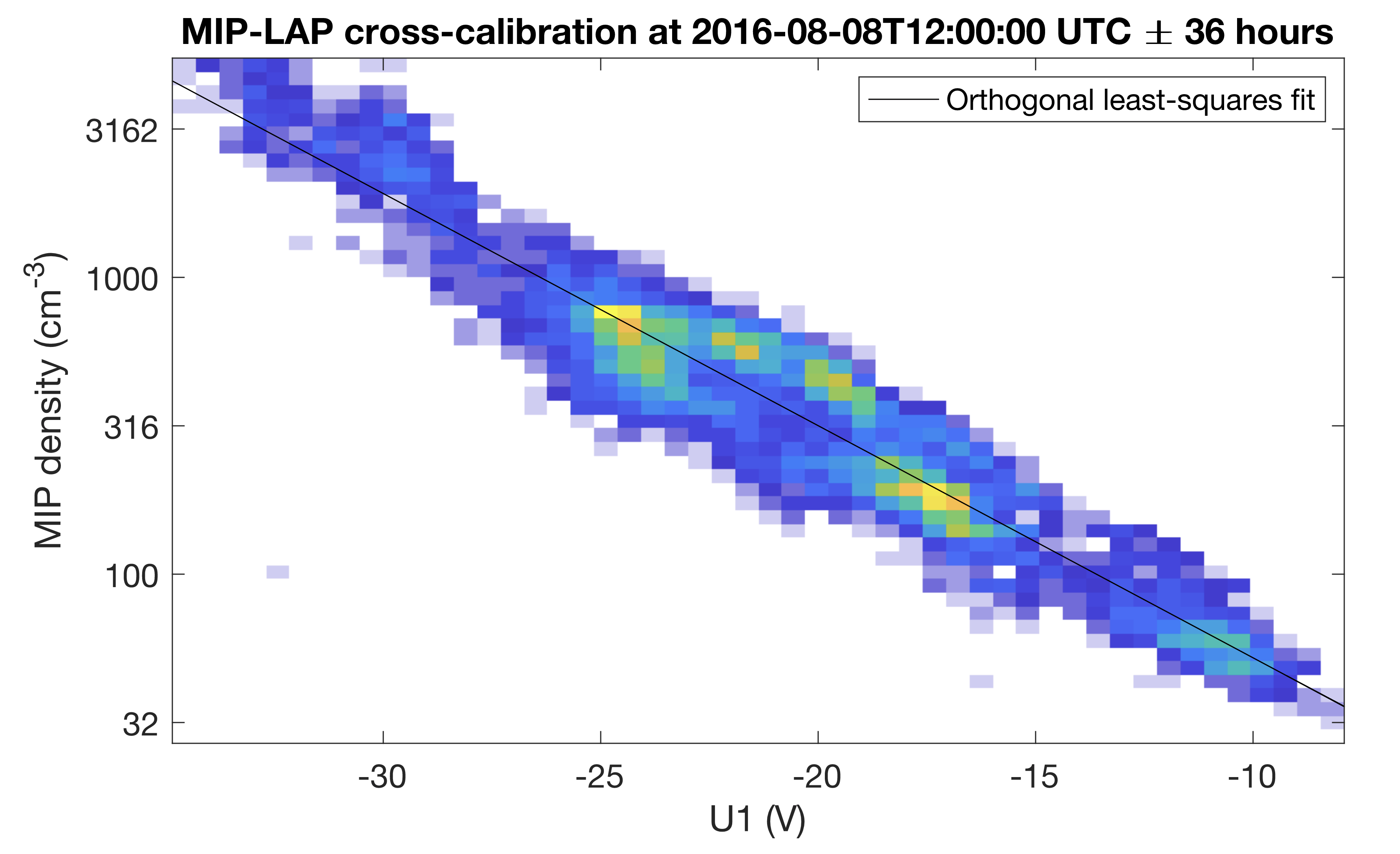}
    \caption{An example of the sliding cross-calibration window. 2D histogram of MIP density vs $U_1$, 50x50 bins coloured by counts. The orthogonal least squares fit (black line) yields an electron temperature of 5.85~eV for $1.053 U_1 = V\ind{S}$, as per Figure~\ref{fig:vzvf_ica}.}
    \label{fig:xcal_ne_u1}
\end{figure}

As several studies \citep{heritier_vertical_2017,Breuillard2019xcal,johansson_charging_2020} show, and as is visible in Figure~\ref{fig:mipne_lapu1}, there is a strong relation between the logarithm of the electron density and the spacecraft potential proxy $U_1$. This relation is even more evident over shorter time windows, shown in Figure~\ref{fig:xcal_ne_u1}, as the photoemission current, which is included in $C$ in Equation~\ref{eq:Vs}, can be assumed to be constant. As the orbital parameters are not uniform over the entire mission, some branching is evident in the mission overview in Figure~\ref{fig:mipne_lapu1} at the lowest densities, which Rosetta only visited on a selection of heliocentric distances, and thus, specific ranges of photoemission current.

The spacecraft potential is, as such, clearly a good proxy for the electron density, and we relate the two in a similar fashion as \citet{Breuillard2019xcal} by rearranging Equation~\ref{eq:Vs}:
\begin{equation} \label{eq:np1p2u1}
    n = P_2 \exp \left( - \frac{U_1}{P_1}\right),
\end{equation}
where $P_1$ and $P_2$ are both constants over the time interval. $P_1$ is defined as $P_1 =\alpha k_B T\ind{e} / e$, with the definition of $\alpha$ from Section~\ref{sec:scpot}.

The cross-calibration of electron density to simultaneous $U_1$ estimates was performed with a 3-day window that is stepped in 1-day steps over the entire mission according to Equation~\ref{eq:np1p2u1}, with some outlier removal, as specified in RO-IRFU-LAP-XCAL\footnote{\label{foot:amdaxcalreport}\url{http://psa.esac.esa.int/pub/mirror/INTERNATIONAL-ROSETTA-MISSION/RPCLAP/RO-C-RPCLAP-5-PRL-DERIV2-V1.0/DOCUMENT/RO-IRFU-LAP-XCAL.PDF}}. This rather long calibration window ensures a physical interpretation of each fit, allows us to bridge longer gaps in MIP data, and generate a continuous and mission-wide electron density estimate, at the time resolution of the LAP $U_1$ estimate. The dynamical range is also much improved both above and below the sensitivity range of MIP regardless of its operational mode. Temporal solar EUV events, or rapid and significant fluctuations in electron temperature will not be correctly captured by the the model, but the general trend in the electron density data will still be valid.

\begin{figure}
\centering
    \includegraphics[width=0.9\columnwidth]{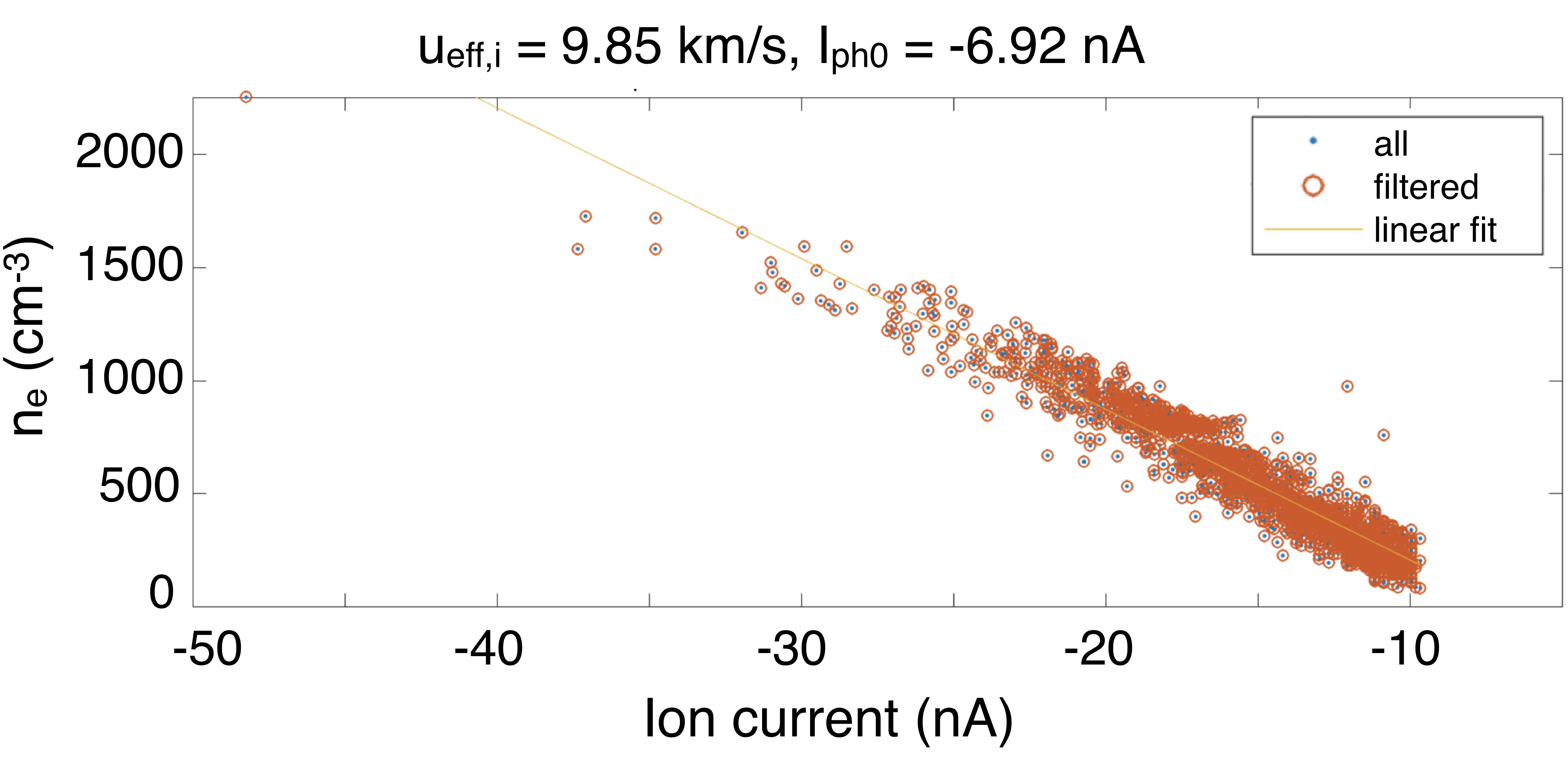}
    \caption{An example of the sliding cross-calibration window for ion current: MIP density vs $I\ind{-}$ for three hours around 2016-05-23T23:30:00. A linear orthogonal least squares fit yields an effective ion speed of 9.85~km/s and an estimate for $I\ind{ph0}$ of -6.92~nA.}
    \label{fig:xcal_ioni_vs_ne}
\end{figure}

\subsection{Cross-calibration of ion current to electron density}
\label{sec:NEL_I}

In a similar, but more straight-forward fashion, one can relate the ion current to the ion density, and by argument of quasi-neutrality, the electron density (under the assumption that charged cometary dust does not contribute much to the overall current balance). At the bias potentials considered ($V\ind{b} \leq -15$~V), the Langmuir probe is repelling electrons, such that the current contribution from plasma electrons can be assumed to be negligible, including secondary emission from electron impact. Whenever cometary cold ions are present in the ICA data, they appear supersonic, i.e.\ the thermal speed of ions is much lower than their flow speed \citep{Bergman_ion_bulk2021}. This is supported also by the identification of a cometary ion wake behind the spacecraft as discussed in \citet{odelstad_ion_2018}. It can be shown~\citep{sagalyn_measurement_1963,fahleson_theory_1967} that for a supersonic flow of ions in a plasma (a cold ion beam), the current to a sunlit probe attracting ions is given by

\begin{equation}\label{eq:Ii}
    I\ind{-} = I\ind{ph0} - n\pi r\ind{p}^2q\ind{i}\sqrt{\frac{2E\ind{i}}{m_i}}\left(1-\frac{q\ind{i}V\ind{p}}{E\ind{i}}\right),
\end{equation}
where $I\ind{ph0}$ is the photosaturation current of the probe, defined to be negative, $r\ind{p}$ is the radius of the probe, $E\ind{i}$ is the energy of ions of charge $q\ind{i}$ and mass $m\ind{i}$, and $V_p$ is the absolute potential of the probe according to $V\ind{p} = V\ind{b} + V\ind{S}$. In this cold ion limit we have neglected the thermal velocity component, and we define the effective ion speed $u\ind{eff,i}$ such that $E\ind{i} = 0.5 m\ind{i}u\ind{eff,i}^2$. For a warm flowing ion plasma, $E\ind{i}$ may instead be interpreted as a typical ion energy, combining the thermal and ram energy. A factor of $4/\pi$ then appears inside the square root term in Equation~(\ref{eq:Ii}) \citep{lindqvist_plasma_1994}, but the linearity of the current-voltage relation remains for all attractive probe potentials \citep{mott-smith_theory_1926,fahleson_theory_1967}.

If we assume that the plasma density varies much faster than the ion energy and the EUV flux (and therefore, $I\ind{ph0}$), and we ignore small perturbations arising from probe potential variations, we can expect a linear relation between $I\ind{-}$ and the electron density.

This is shown in an example cross-calibration window in Figure~\ref{fig:xcal_ioni_vs_ne}, where the x-intersect yields the photosaturation current. In a similar fashion as in Section~\ref{sec:NEL_V}, we perform a linear orthogonal least square fit of coinciding $I\ind{-}$ measurements and MIP electron density estimates over a window spanning three hours, which is stepped with one hour over the entire cometary mission. If there are several $I\ind{-}$ measurements, during a MIP density sampling interval, an average is taken. Measurements of $I\ind{-}$ for a shadowed probe are analysed separately, assuming $I\ind{-}(n=0) = 0$. Some outliers are removed according to RO-IRFU-LAP-XCAL\cref{foot:amdaxcalreport}. 

During periods with low electron density and few coinciding MIP and LAP data points (before 2015-01-01 and during the so-called night side excursion around 2016-04-01), where no electron density estimates would otherwise be produced, the calibration instead considers a combined dataset of MIP and LAP sweep densities (obtained from fits as in the example in Figure~\ref{fig:sweepexample}) for the cross-calibration. In this case, the cross-calibration is applied over a larger calibration window (15-day window, 5-day step size) to improve the fitting performance. As the spacecraft potential is low or positive during these periods, the reliability of the LAP sweep electron density estimates is considered to be the best available, although still sensitive to electron temperature variations.

 \section{Results}
 \label{sec:results}
\begin{figure}
    \includegraphics[width=1.0\columnwidth]{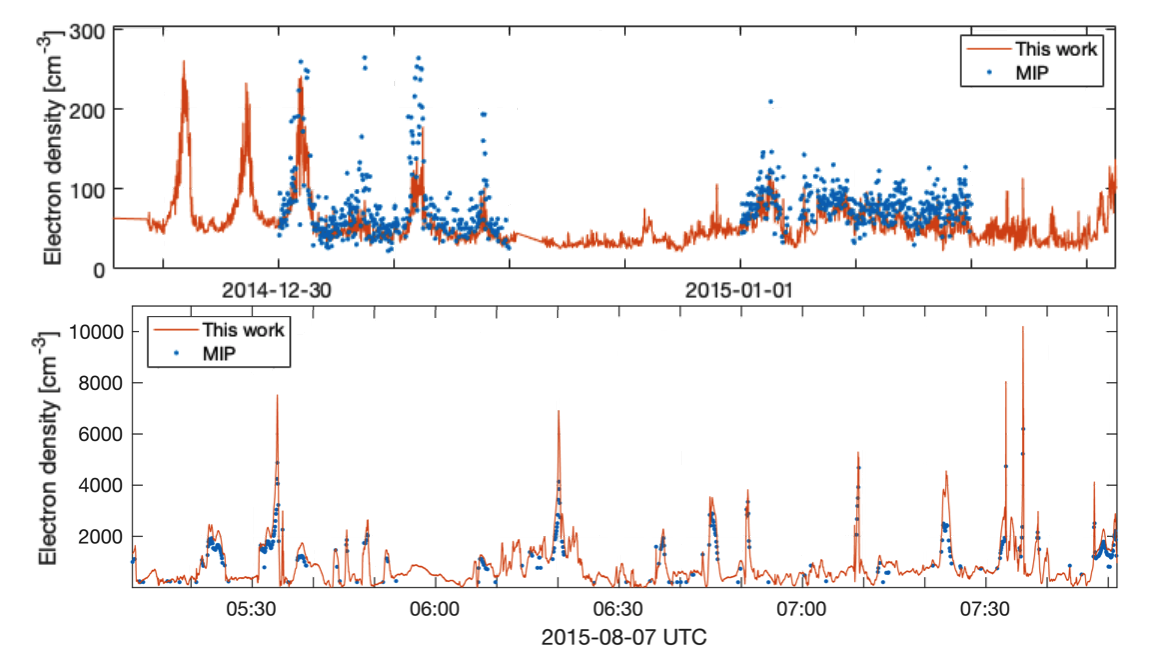}
    \caption{Two examples that illustrates the performance of the cross-calibration. \textbf{Top:} New Years Eve 2014, when the MIP instrument (blue dots) was occasionally in a measurement mode that is not sensitive to low densities, but LAP was continuously measuring spacecraft potential, and recovers the density via cross-calibration to spacecraft potential (red line). The errors in the cross-calibration are dominated by uncertainties in MIP, and are typically 25 percent in both datasets. \textbf{Bottom:} The time resolution of the electron density dataset is also improved, here via a $\sim$60~Hz ion current cross-calibration as described in Section~\ref{sec:NEL_I}.}
    \label{fig:xcal_nednelexamples}
\end{figure}

\begin{figure*}
    \includegraphics[width=0.998\textwidth]{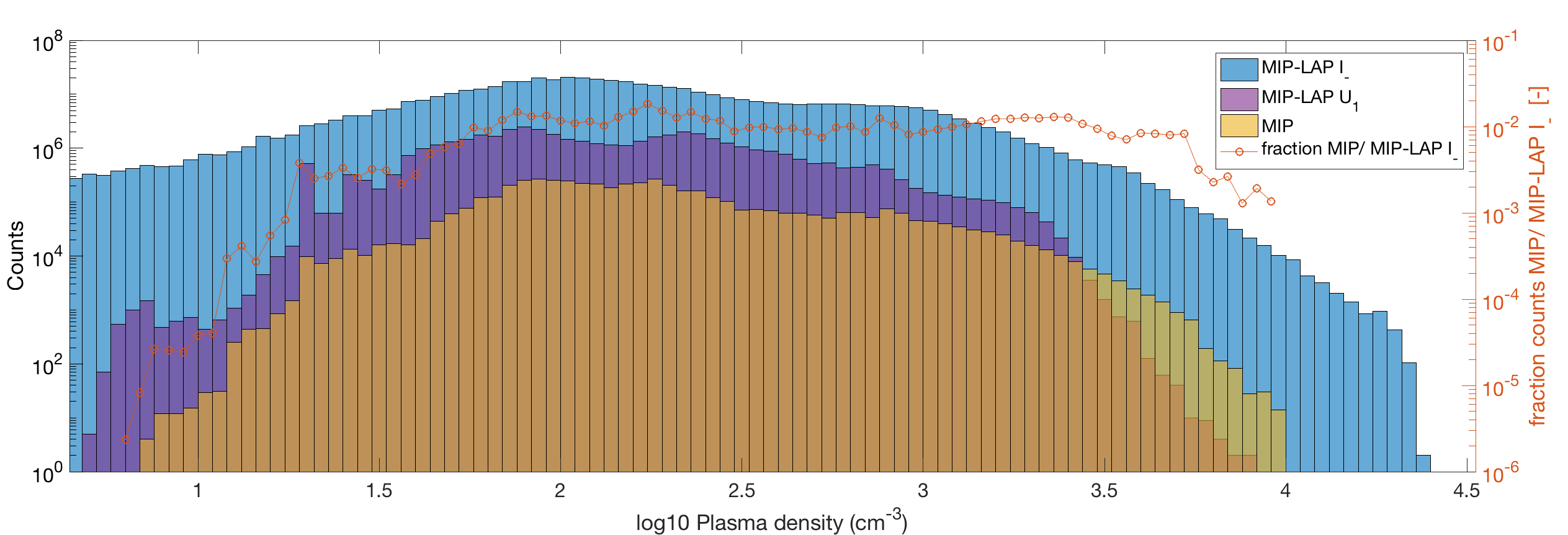}
    \caption{Mission-wide histograms of electron density from the cross-calibration of LAP ion current to MIP in blue bars, of spacecraft potential to MIP (purple) and the MIP densities in yellow bars. Also plotted on the axis to the right is the fraction of MIP counts to MIP-LAP $I_-$ counts in each density bin. As the LAP current measurement has finite resolution, fractional errors exceed unity below 5~cm$^{-3}$ even for a perfect calibration of the ion current to density, and are therefore not included.}
    \label{fig:hist_neli_nel_mip}
\end{figure*}

\begin{figure*}
\centering
    \includegraphics[width=1.0\textwidth]{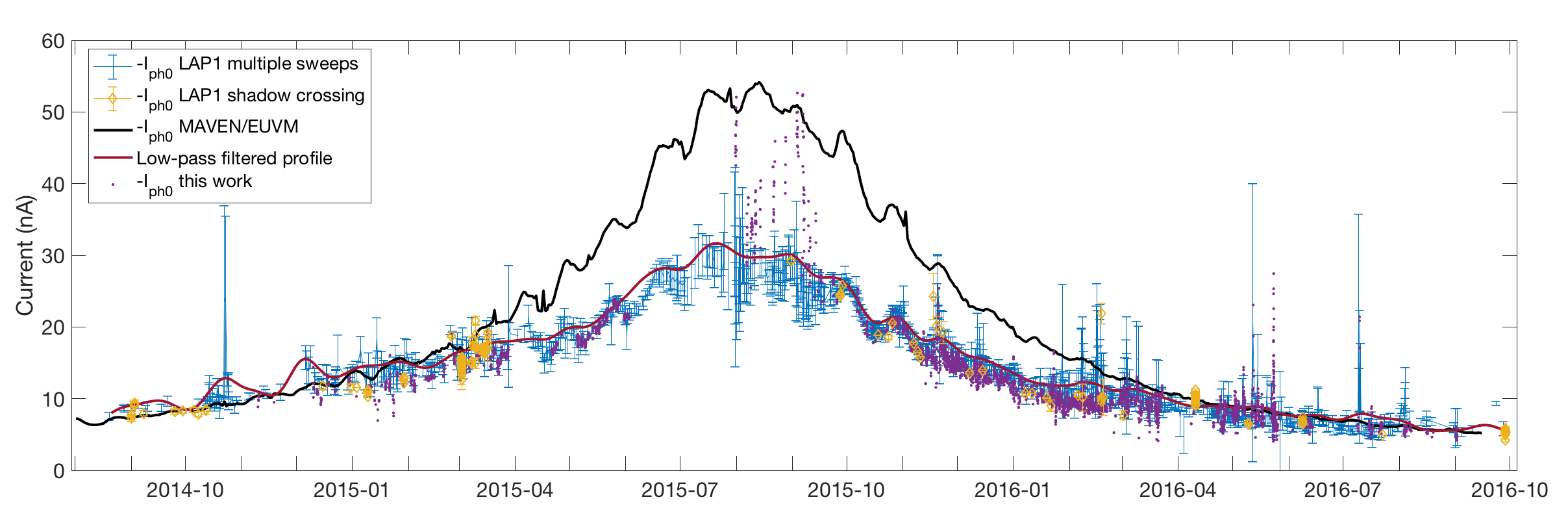}
    \caption{Photosaturation current estimates during the cometary mission from the cross-calibration of ion current to MIP (purple dots) with an associated error bar from the fit of less than five percent, three different photosaturation current estimates are as described in \citet{johansson_rosetta_2017}, where blue and yellow are two out of three independent estimates from LAP with associated error bars, and the black line is a(flare-removed) photosaturation estimate as measured from an EUV monitor at Mars orbit. Also plotted, a low-pass ($1/f<9.5$~days) filtered profile of all photosaturation estimates from LAP (red line).}
    \label{fig:iph0_all}
\end{figure*}

The performance of the ion current cross-calibration procedure can be seen in Figure~\ref{fig:xcal_nednelexamples}. Here, we recover densities in periods where MIP does not produce electron densities, and (bottom) the temporal resolution has increased dramatically, but is well in agreement with MIP density estimates whenever available.

 For comparison and as a mission overview, we plot a histogram of the resulting cross-calibrated plasma density datasets, as well as the MIP density in Figure~\ref{fig:hist_neli_nel_mip}. The cross-calibrated datasets cover a wider dynamic range of densities, at least for the ion current, and for density ranges fully within the range of the MIP sensitivity, we see a near constant ratio between the two datasets, as expected for two comparable measures with unequal temporal resolution. Also as expected, the ratio drops at the lower edge of the MIP sensitivity range. As noted at the end of Section~\ref{sec:NEL_V}, the ion current, the MIP density and the spacecraft potential are all measures at slightly different spatial scales. And even though there is no obvious reason why the density estimates from the spacecraft potential method has an upper limit, estimates above 3000 cm$^{-3}$ on this (32~m) scale seems rare with this method, and is instead most common on Langmuir probe current scales. A feature indicative perhaps of the sizes of the plasma structures that move past the Rosetta spacecraft.

The physical interpretation of the cross-calibration coefficients yields an estimate for the photosaturation current $I\ind{ph0}$ from the x-intersect of the fit in Figure~\ref{fig:xcal_ioni_vs_ne} as $I\ind{-}(n=0) = I\ind{ph0}$. This new estimate of $I\ind{ph0}$ agrees very well with other methods of obtaining the photoemission current published in \citet{johansson_rosetta_2017}, as shown in Figure~\ref{fig:iph0_all}.

\begin{figure*}
\centering
    \includegraphics[width=1.0\textwidth]{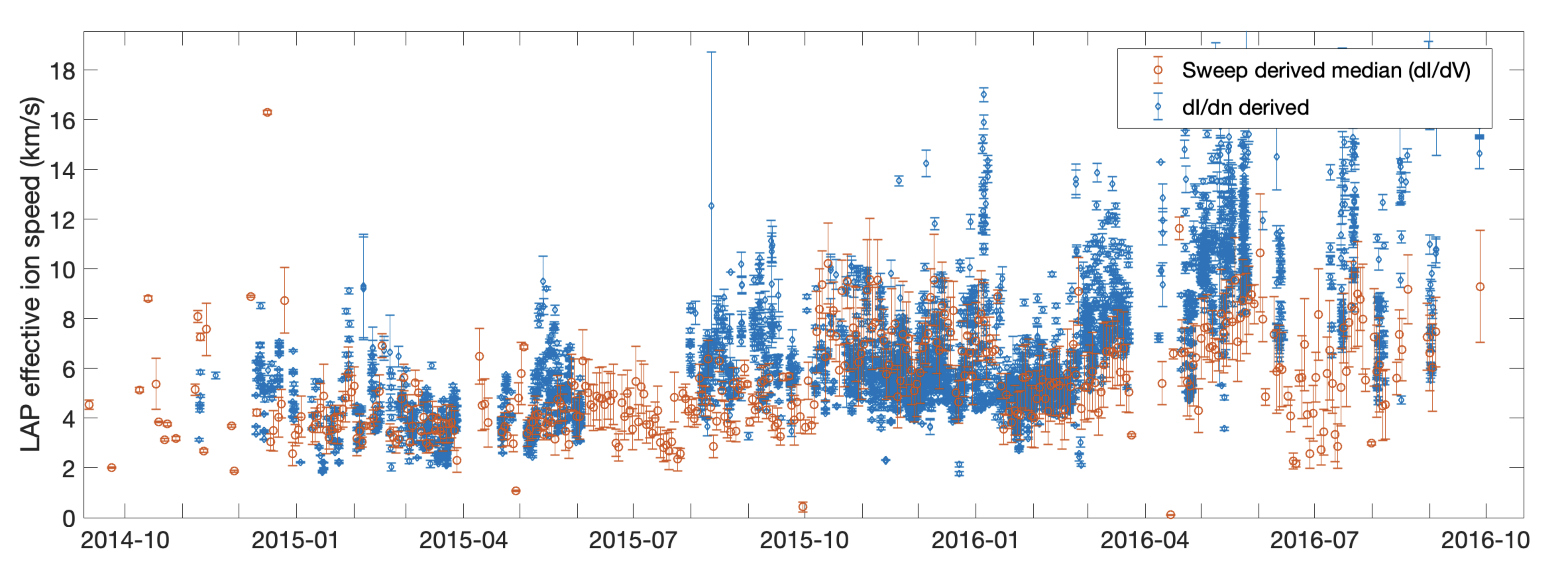}
    \caption{Effective ion speed estimates from the slope of the ion current, ${\rm d}I/{\rm d}V\ind{b}$, from Langmuir Probe sweeps (median of 19 hour bins plotted in red) and from ${\rm d}I/{\rm d}n$ using three hour long $\sim$60~Hz ion current cross-calibration windows (blue). Both methods are using coinciding plasma density estimates from MIP.}
    \label{fig:xcal_v}
\end{figure*}

For the slope coefficient of the ion current cross-calibration fit (${\rm d}I/{\rm d}n$), we can solve for the effective ion speed via Equation~\ref{eq:Ii} assuming the spacecraft potential fluctuation is small compared to $V\ind{p}$ and that ions are singularly charged with a mass of 19~amu~\citep{heritier_ion_2017}. The resulting speed is plotted in Figure~\ref{fig:xcal_v} and compared with a sweep derived ion speed estimate (${\rm d}I/{\rm d}V$), where we evaluate the slope of the ion current ${\rm d}I/dV\ind{b}$ using a linear orthogonal least square fit (see Figure~\ref{fig:sweepexample}), as well as a coinciding MIP density measurement, as described in detail in \citet{vigren_effective_2017}. Random errors are estimated from the uncertainty in the least square fits, assuming the errors are normally distributed, and is dominated by MIP sweep frequency discretisation. As the number of MIP and LAP measurement points are much greater in the cross-calibration window than in the sweep estimates, we try to reduce the random errors in the sweep estimates by binning the data in 500 bins of equal length (19~hour) over the entire mission.

\section{Discussion} \label{sec:discussion}

Before discussing the implications of the cross-calibration, where we focus on the ion-current cross-calibration, we shall first discuss error sources.

\subsection{Errors}

In the cross-calibrated densities, the random errors associated with the uncertainty of the associated LAP measurement is very small, typically 5~cm$^{-3}$, assuming an ion velocity of 5~kms$^{-1}$. However, the standard error in the calibration is dominated by the uncertainty of the associated MIP measurement (typically 25~percent), but is also influenced on the fluctuation of the EUV flux to the probe (if sunlit) or the spacecraft during the calibration interval (typically less than 2~nA and much less than 10~percent). As the final fractional error (assuming the errors are normally distributed) is the sum of the squares of the fractional errors, the error margins of MIP and the cross-calibrated densities is more or less identical, and for the comparison in Figure~\ref{fig:xcal_nednelexamples} we plot all datasets without the $\approx$ 25~percent fractional uncertainties.

Systematic errors arise from the limited validity of the underlying assumption that the ion density at the probe position is equal to the electron density MIP measures, and at least proportional to the average electron current to the spacecraft, which is an error we cannot easily quantify. If this assumption is not correct, the slope in the fit, and therefore the effective velocity of the ions we estimate from the ion current cross-calibration should diverge from the true value by some factor equal to the difference in ion and electron densities during the calibration interval. Such effects include sheath effects from the spacecraft potential for ions, and may require further investigation with spacecraft-plasma interaction simulations. However, we can test the assumption by comparing the effective velocity estimate to ICA ion measurements, as we will do in subsection~\ref{sec:disc_effion}

If the ion velocity (for the ion current cross calibration) or the electron temperature of the thermal electron population that dictates the spacecraft potential dependence on electron density \citep{johansson_charging_2020} is not constant during the cross-calibration interval, errors would grow, and the resulting fit would be poor. Effective ion velocities and photoemission currents resulting from fits with correlation coefficient less than 0.7 have therefore been removed.

It should also be noted that the ion current to a Langmuir probe from ions with a superposition of velocity distributions is the superposition of the current from each velocity distribution. Although a Maxwellian simplifies the interpretation of the LAP effective ion speed, it is not dependent on a Maxwellian assumption. The equations that govern the cross-calibration and the ion speed estimates rely on Orbital Motion Limited (OML) theory, i.e. on the assumption that the Langmuir probe size (5~cm) is much smaller than the Debye length, which is generally true for the entire mission \citep{Gilet2019cold, wattieaux_plasma_2020}. If the Debye length is at any point smaller than the probe size, in very cold and dense plasmas, we note that OML theory yields an upper limit on the collected current. In such cases, we would expect to see more points above the linear trend in Figure~\ref{fig:xcal_ioni_vs_ne} when the density is high, which we generally do not. Similarly, it would yield a significant underestimation of the photoemission when the density is high, but this is not observed in Figure~\ref{fig:iph0_all}. Additionally, if the thermal velocity is the dominant component of the velocity, in contrast to the findings in \citet{odelstad_ion_2018} and \citet{Bergman_ion_bulk2021}, our cold ion approximation in Equation~\ref{eq:Ii} would lead to an overestimation of $u\ind{eff,i}$ by 13~percent.

\subsection{EUV intensity}

We are able to resolve the solar sidereal rotational period of $\sim$26~days in our photosaturation current from the cross-calibration plotted in Figure~\ref{fig:iph0_all}, and also obtain a good agreement with other photosaturation estimates from LAP (of which the shadow-crossing method is believed to be the most accurate). Therefore, we find this result to be another confirmation of the results published in \citet{johansson_rosetta_2017} as well as of the physical model behind the cross-calibration.

We also note that if there is a significant amount of higher energy ($>20$~eV) electrons, such as the population giving rise to electron-impact ionisation in the cometary plasma, this population would also be able to excite and emit secondary electrons from the probe surface. For Cassini in the Saturn magnetosphere, \citet{garnier_detection_2012} found such secondary emission to be a significant contributor to the Langmuir probe current at negative bias voltage. The source population in this case was identified as electrons with energy in the approximate range 250-450~eV. As significant electron fluxes in this energy range have been detected also by Rosetta \citep[e.g.\ ][]{clark_suprathermal_2015,Myllys_plasma_prop2019} one may ask if such currents appreciably contaminate our photoemission estimates. If the secondary yield is higher than 1, i.e.\ more than one electron emitted per incoming high energy electron, then this would constitute an additive offset in all LAP measured $I\ind{ph0}$ estimates. However, the shadow crossing method is insensitive to this secondary emission, except in the highly unlikely case of the electron flux being parallel to the direction from the Sun. The good agreement between the shadow crossing method with other LAP estimates, whenever available, suggests that secondary electron emission due to impact of high energy electrons generally has little to no effect on our results.

\subsection{Effective ion speed} \label{sec:disc_effion}

The ion speed estimate from the cross-calibration agrees generally very well with the sweep derived ion speed in Figure~\ref{fig:xcal_v}, and again confirms the validity of the physical model in the cross-calibration. They are also very much in line with the average flow velocities as reported by ICA~\citep{nilsson_average2020mnras}, also during the events investigated in detail by \citet{Bergman_ion_bulk2021} as well as previously published LAP estimates \citep{vigren_effective_2017,odelstad_ion_2018}. Since the two velocity estimates in Figure~\ref{fig:xcal_v} arise from two different methods (${\rm d}I/{\rm d}V\ind{b}$ vs ${\rm d}I/{\rm d}n$), where the internal instrumental offsets are very different (as one involves stepping voltages with the LAP bias circuitry), we conclude that we have a very good grasp of instrument calibration and (as all $I\ind{ph0}$ estimates agree), a near-zero contamination on LAP1.

\begin{figure}
    \includegraphics[width=1.0\columnwidth]{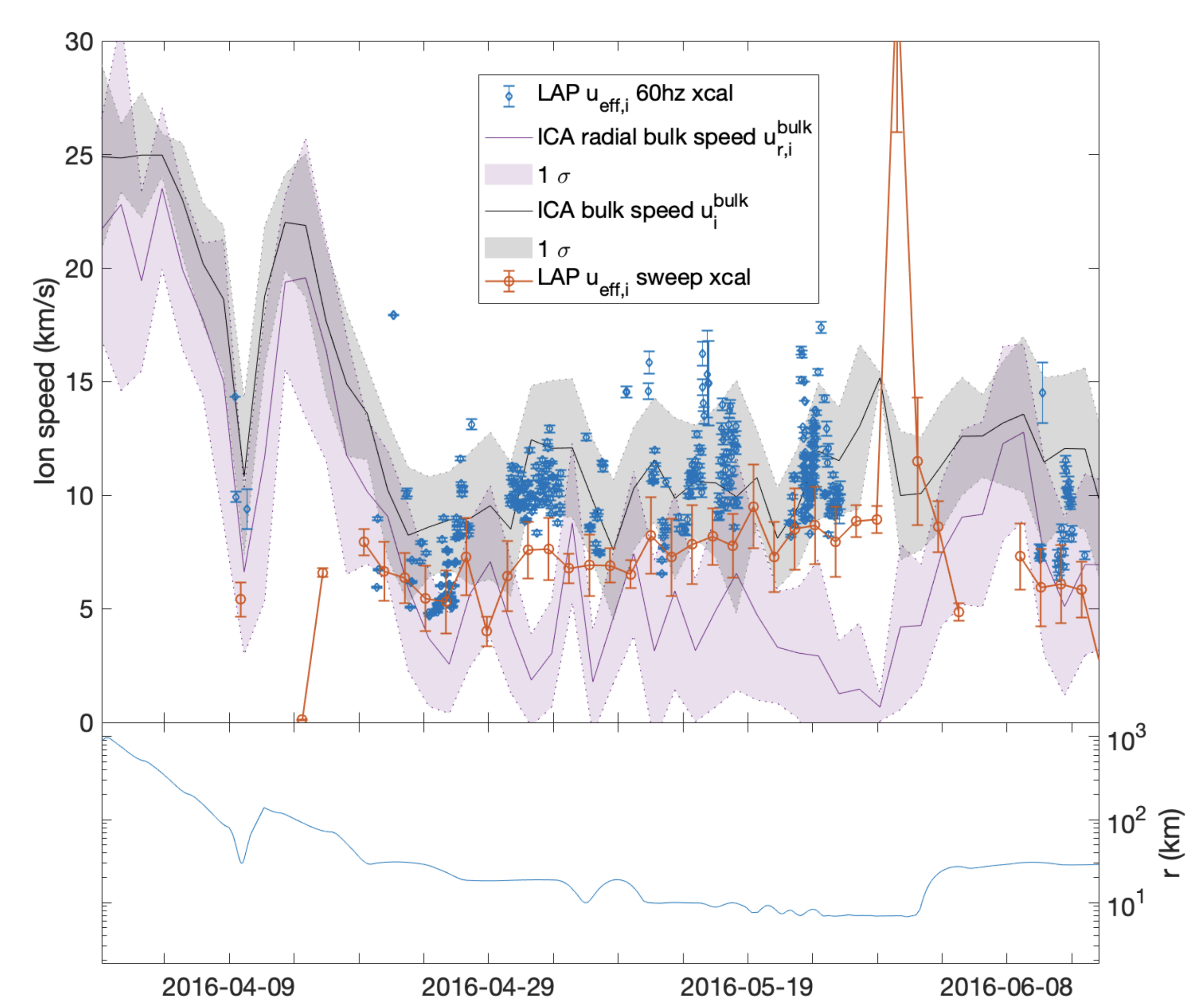}
    \caption{\textbf{Top:} Median of 19~hour bins of ICA H$\ind{2}$O$\super{+} $ion bulk velocities in terms of magnitude (black line) and purely radial component (purple line) with corresponding  $\pm 1 \sigma$ spread (shaded) from the standard deviation of the mean, as well as the two LAP effective ion speed estimates from the cross-calibration in Section~\ref{sec:NEL_I} (blue line) and 19~hour binned medians from sweeps (red line) vs time for a selected time period where the radial distance was rapidly varied. \textbf{Bottom:} radial distance vs time for the same time period. At $\sim$ 1000~km to 30~km the radial component of the ion velocity seems to dominate, and we have good agreement for all estimates, when available. Even closer to the comet nucleus the radial component decreases and perpendicular components grow, and as expected, both LAP effective ion speed estimates increase as the magnitude of the velocity increases.}
    \label{fig:ICA_vr_LAP_201604}
\end{figure}

However, for the entire cometary mission, the ion speed estimate is much larger than the neutral velocity of 0.4-1~kms$^{-1}$ \citep{hansen_evolution_2016} and thus speaks against the assumption made in previous ionospheric models~\citep{heritier_vertical_2017,heritier_plasma2018,galand_ionospheric_2016}, at large heliocentric distances, that ions are flowing with the neutral gas speed. This assumption is not strictly necessary for such ionospheric model density result, and we note that to reconcile our ion speed estimate with these models, either the radial velocity of ions is not the dominant component of the ion flow, or there have to be some mechanism that increase ionisation between Rosetta and the comet nucleus.

As LAP can not resolve ion density independently, we note that any spacecraft sheath related effects which would alter the ion density at the probe location would affect both LAP estimates equally. Validating the ion density with ICA over the entire mission is not currently feasible
, mainly due to the limited and distorted field of view \citep{bergman_influence_2019,bergman_influence_debye_2020}, such that the ion density estimate would differ from the expectation value at any time, depending on geometry and ion flow direction. However, this has a very limited effect on the ICA ion speed estimate \citep{Bergman_ion_bulk2021}. Except for an ion beam that is invariably outside of the ICA instrument field of view during the entire Rosetta mission, we see no reason why the cometary bulk ion speed should be systematically incorrect after a simple correction for the spacecraft potential.

Therefore, to further cross-validate our result, we plot the two LAP estimates together with the magnitude and the radial component of the ICA $<60$~eV H$\ind{2}$O$\super{+}$ (or H$\ind{3}$O$\super{+}$) bulk drift velocity  ($u_i^{bulk}$ and ${u_{r,i}^{bulk}}$, respectively) from \citet{nilsson_average2020mnras} in Figure~\ref{fig:ICA_vr_LAP_201604} for an interval near the tail-side excursion. The estimates are not perfectly equivalent but should still correlate well with the magnitude of the ICA estimate, when available.

Here, we find (within error bars) that the radial velocity is dominant and positive at large cometocentric distances (from $\sim$~ 30 to 1000~km), and increasing with radial distance, which was also noted in the beginning of this interval in \citet{Behar_root_2018}. Also, as expected, both LAP effective ion speed estimates correlate and agree within error bars with the magnitude of the ICA H$\ind{2}$O$\super{+} $ bulk velocity when available, pointing to the existence of a radial electric field such as an ambipolar electric field \citep{madanian_suprathermal_2016,vigren_1d_2017,bercic_cometary_2018,deca_building2019prl}. As the ambipolar electric field is proportional to the electron pressure gradient, a radial expansion is consistent with increasing radial ion velocities with cometocentric distance. Moreover, an ambipolar electric field would accelerate electrons falling inward and, as more electrons reach electron-impact ionisation energies, provide increased ionisation between Rosetta and the comet nucleus, as postulated earlier in this section.

As Rosetta moves below 30~km, the radial velocity component decreases, but perpendicular components grow and start to dominate. We suspect that this could again be an effect of an ambipolar electric field as the comet activity is inhomogeneous and bursty by nature, which would create strong non-radial electron-pressure gradients, and therefore, form an ambipolar electric field with non-radial components. It seems reasonable to expect this field to be strongest near the nucleus, where gas density gradients are great. Also, ions accelerated by a non-radial field close to the comet would also appear radial for a faraway observer with a limited viewing angle resolution. This hypothesis is highly speculative, but could perhaps be tested in the future using simulations similar to \citet{deca_building2019prl} and \citet{Divin_kinetic2020}, but resolving the nucleus and allowing for a more realistic outgassing.

\begin{figure}
    \includegraphics[width=1.0\columnwidth]{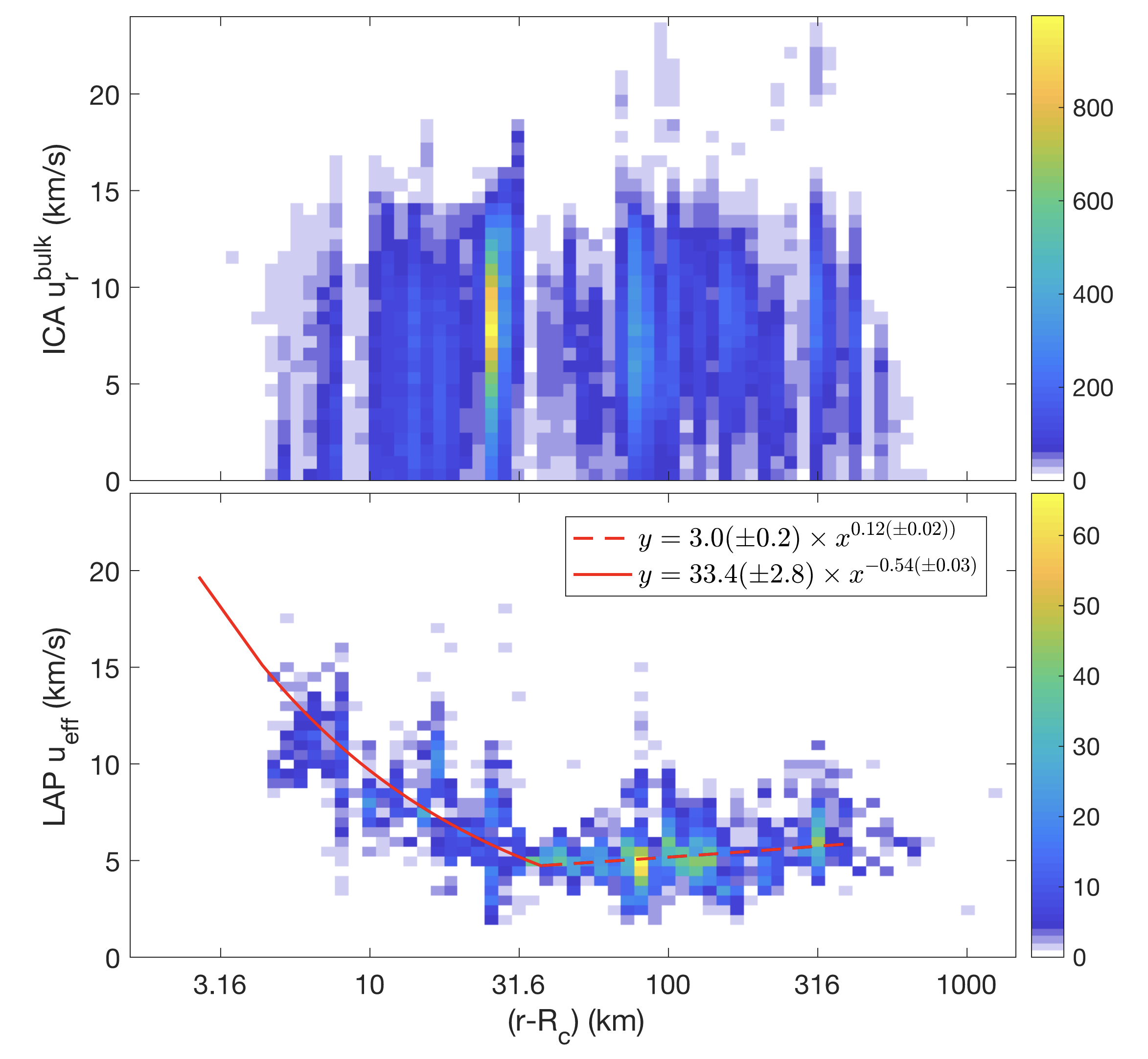}
    \caption{2-D histograms of ion speed vs distance to comet surface coloured by counts, with a common x-axis. \textbf{Top:} Radial bulk speed of water ions from ICA, $u^{bulk}_{r,i}$ in 90x40 bins. \textbf{Bottom:} The union of the two LAP effective flow speed $u\ind{eff,i}$ estimates in 90x40 bins and two least-square regression fit (red solid or dashed line) for two selected intervals, with errors in the coefficients estimated from the least-square regression.}
    \label{fig:ICA_vr_LAP_hists}
\end{figure}

This trend seems also generally true for the entire mission, as plotted in Figure~\ref{fig:ICA_vr_LAP_hists}, where the combined dataset of the two LAP effective ion speeds, $u\ind{eff,i}$, shows large (presumably non-radial) velocities below 30~km from the comet surface. We also find and an acceleration with radial distance beyond 30~km distance, as reported by \citep{bercic_cometary_2018} using ICA measurements.

Nevertheless, all measured estimates suggest that the ions are moving much faster than the neutrals over the entire mission. The discrepancy between $u\ind{eff,i}$ and ${u_{r,i}^{bulk}}$ outside 30~km from the comet surface, although there is overlap, could perhaps be attributed to that the ion density is enhanced at the probe position. If so, the assumption of quasineutrality via $n\ind{e} = n\ind{i}$, is incorrect, perhaps by significant electron depletion by dust \citep{morooka_dusty_2011}, and would need further investigation. There might also be field of view effects on the (mostly nadir pointing) ICA instrument, which warrants further investigation, and is why we, in line with \citet{nilsson_average2020mnras}, limit our comparison to the radial velocity component in ICA.

If there was a strong bias against ions at lower energies in both ICA and LAP, for instance if the spacecraft potential more easily deflects low energy ions away from detection surfaces, we would expect there to be a strong dependence on the drift energies on spacecraft potential shown in Figure~5 in \citet{nilsson_average2020mnras}, but we find none. Additionally, as shown in a recent case study by \citet{Bergman_ion_bulk2021}, there were no indication that there is a significant low-energy cometary ion population present, as detailed spacecraft-plasma interaction simulations of such a environments were incompatible with ICA measurements. Moreover, the instrument field of view increases for low energy ions on a negatively charged spacecraft \citep{bergman_influence_2019}, but some uncertainties still remain regarding the geometric factor at low ion energies \citep{Bergman_ion_bulk2021}. Still, the two LAP estimates, as well as the ICA ion velocity shows elevated ion velocities, for which at least the radial component increases with radial distance. This is consistent with a radial ambipolar electric field that has been predicted to be capable of accelerating the cold new-born ions from the neutral speed to the speed we observe \citep{vigren_1d_2017}.

\section{Conclusions}

We have devised and verified two methods to recover a mission-wide plasma density dataset from the LAP spacecraft potential estimates and the LAP ion current by cross-calibrating the estimate to MIP density whenever available. As a result, we improve the dynamic range as well as the temporal resolution of the RPC plasma density dataset up to a factor of 240 and facilitate plasma analysis at much shorter timescales. The dataset has been made available on the PSA~\citep{RosettaPSA}, and at the CDPP on AMDA\footnote{\url{http://amda.irap.omp.eu}}. The spacecraft potential and the effective ion speeds resulting from the MIP-LAP ion current cross-calibration model have been successfully cross-validated with ICA, an instrument with a fundamentally different measurement principle.

The physical model that enables the cross-calibration allows for an almost continuous (3~hour cadence) estimate of the effective ion speed and, when the probe is sunlit, the photosaturation current. The latter we find to be well in agreement with independent methods in published studies \citep{johansson_rosetta_2017} and provides support for the conclusions drawn therein regarding attenuation of the EUV.

The ion speed estimates are found to be large $\sim$ 5~kms$^{-1}$ and mostly radial in altitudes above $\sim$30~km, which is in line with previously published LAP-derived ion speeds \citep{odelstad_ion_2018,vigren_effective_2017}, and with recent ICA estimates of the ion bulk velocity \citep{nilsson_average2020mnras, Bergman_ion_bulk2021} of H$\ind{2}$O$\super{+}$, but in disagreement with the assumption that the ions flow with the speed of the neutrals. This assumption has been made in several modelling works that target plasma densities, and in an average sense successfully reproduce observations at low activity \citep{vigren_model-observation_2016,galand_ionospheric_2016,heritier_ion_2017,heritier_vertical_2017,heritier_plasma2018}.
As a faster radial ion flow would decrease these model density estimates at the Rosetta position, some other error, such as a process which increases the rate of ionisation must also be present. However, further investigation is needed to confirm that the inferred bulk speeds are representative of the actual drift speed.

The elevated velocities in and of itself points to an electric field present throughout the entire cometary mission, capable of accelerating ions and increasing ionisation via electron impact ionisation between Rosetta and the comet nucleus. Other candidates for the elevated ion speeds (above the neutral speed the ions are born at) includes the wave processes already detected at the comet \citep{andre_lower_2017,ruhunusiri_turbulence_67P2020JGR,karlsson_rosetta_2017}.

\begin{acknowledgements}
      \emph{Rosetta} is an ESA mission with contributions from its member states and NASA. This work would not have been possible without the collective efforts over a quarter of a century of all involved in the project and the RPC. This research was funded by the Swedish National Space Agency under grant Dnr 168/15. The cross-calibration of LAP and MIP data was supported by ESA as part of the Rosetta Extended Archive activities, under contract 4000118957/16/ES/JD. Work at LPC2E is also supported by CNES. We acknowledge the staff of Centre de Données de la Physique des Plasmas (CDPP) for the use of Automated Multi-Dataset Analysis (AMDA). We also acknowledge the ESA Planetary Science Archive for archiving and reviewing the LAP data.
\end{acknowledgements}

\bibliographystyle{aa}
\bibliography{References_all}

\begin{thebibliography}{50}
\expandafter\ifx\csname natexlab\endcsname\relax\def\natexlab#1{#1}\fi

\bibitem[{André {et~al.}(2017)André, Odelstad, Graham, Eriksson, Karlsson,
  Wieser, Vigren, Norgren, Johansson, Henri, Rubin, \&
  Richter}]{andre_lower_2017}
André, M., Odelstad, E., Graham, D.~B., {et~al.} 2017, \mnras, 469, S29

\bibitem[{{Behar} {et~al.}(2018){Behar}, {Nilsson}, {Henri},
  {Ber{\v{c}}i{\v{c}}}, {Nicolaou}, {Stenberg Wieser}, {Wieser}, {Tabone},
  {Saillenfest}, \& {Goetz}}]{Behar_root_2018}
{Behar}, E., {Nilsson}, H., {Henri}, P., {et~al.} 2018, \aap, 616, A21

\bibitem[{Bergman {et~al.}(2019)Bergman, {Stenberg Wieser}, Wieser, Johansson,
  \& Eriksson}]{bergman_influence_2019}
Bergman, S., {Stenberg Wieser}, G., Wieser, M., Johansson, F.~L., \& Eriksson,
  A. 2019, Journal of Geophysical Research: Space Physics, e2019JA027478

\bibitem[{Bergman {et~al.}(2020)Bergman, {Stenberg Wieser}, Wieser, Johansson,
  \& Eriksson}]{bergman_influence_debye_2020}
Bergman, S., {Stenberg Wieser}, G., Wieser, M., Johansson, F.~L., \& Eriksson,
  A. 2020, Journal of Geophysical Research: Space Physics, 125, e2020JA027870

\bibitem[{Bergman {et~al.}(2021)Bergman, Wieser, Wieser, Johansson, Vigren,
  Nilsson, Nemeth, Eriksson, \& Williamson}]{Bergman_ion_bulk2021}
Bergman, S., Wieser, G.~S., Wieser, M., {et~al.} 2021, \mnras, stab584

\bibitem[{Berčič {et~al.}(2018)Berčič, Behar, Nilsson, Nicolaou, Wieser,
  Wieser, \& Goetz}]{bercic_cometary_2018}
Berčič, L., Behar, E., Nilsson, H., {et~al.} 2018, \aap, 613, A57

\bibitem[{Besse {et~al.}(2018)Besse, Vallat, Barthelemy, Coia, Costa, {De
  Marchi}, Fraga, Grotheer, Heather, Lim, Martinez, Arviset, Barbarisi,
  Docasal, Macfarlane, Rios, Saiz, \& Vallejo}]{PSA_2017}
Besse, S., Vallat, C., Barthelemy, M., {et~al.} 2018, \planss, 150, 131

\bibitem[{Breuillard {et~al.}(2019)Breuillard, {Henri, P.}, {Bucciantini, L.},
  {Volwerk, M.}, {Karlsson, T.}, {Eriksson, A.}, {Johansson, F.}, {Odelstad,
  E.}, {Richter, I.}, {Goetz, C.}, {Valli\`eres, X.}, \& {Hajra,
  R.}}]{Breuillard2019xcal}
Breuillard, H., {Henri, P.}, {Bucciantini, L.}, {et~al.} 2019, \aap, 630, A39

\bibitem[{Carr {et~al.}(2007)Carr, Cupido, Lee, Balogh, Beek, Burch, Dunford,
  Eriksson, Gill, Glassmeier, Goldstein, Lagoutte, Lundin, Lundin, Lybekk,
  Michau, Musmann, {H. Nilsson}, Pollock, Richter, \&
  Trotignon}]{carr_rpc:_2007}
Carr, C., Cupido, E., Lee, C. G.~Y., {et~al.} 2007, Space Sci. Rev., 128, 629

\bibitem[{Clark {et~al.}(2015)Clark, {Broiles, T. W.}, {Burch, J. L.},
  {Collinson, G. A.}, {Cravens, T.}, {Frahm, R. A.}, {Goldstein, J.},
  {Goldstein, R.}, {Mandt, K.}, {Mokashi, P.}, {Samara, M.}, \& {Pollock, C.
  J.}}]{clark_suprathermal_2015}
Clark, G., {Broiles, T. W.}, {Burch, J. L.}, {et~al.} 2015, \aap

\bibitem[{Deca {et~al.}(2019)Deca, Henri, Divin, Eriksson, Galand, Beth,
  Ostaszewski, \& Hor\'anyi}]{deca_building2019prl}
Deca, J., Henri, P., Divin, A., {et~al.} 2019, Phys. Rev. Lett., 123, 055101

\bibitem[{{Divin} {et~al.}(2020){Divin}, {Deca}, {Eriksson}, {Henri},
  {Lapenta}, {Olshevsky}, \& {Markidis}}]{Divin_kinetic2020}
{Divin}, A., {Deca}, J., {Eriksson}, A., {et~al.} 2020, \apjl, 889, L33

\bibitem[{Edberg {et~al.}(2015)Edberg, Eriksson, Odelstad, Henri, Lebreton,
  Gasc, Rubin, Andre, Gill, Johansson, Johansson, Vigren, Wahlund, Carr,
  Cupido, Glassmeier, Goldstein, Koenders, Mandt, Nemeth, Nilsson, Richter,
  Wieser, Szego, \& Volwerk}]{edberg_spatial_2015}
Edberg, N. J.~T., Eriksson, A.~I., Odelstad, E., {et~al.} 2015, Geophys. Res.
  Lett., 42, 4263

\bibitem[{Eriksson {et~al.}(2007)Eriksson, Boström, Gill,
  {\textbackslash}AAhlén, Jansson, Wahlund, André, Mälkki, Holtet, Lybekk,
  Pedersen, Blomberg, \& {the LAP team}}]{eriksson_rpc-lap:_2007}
Eriksson, A.~I., Boström, R., Gill, R., {et~al.} 2007, Space Sci. Rev., 128,
  729

\bibitem[{Eriksson {et~al.}(2017)Eriksson, Engelhardt, André, Bostrōm,
  Edberg, Johansson, Odelstad, Vigren, Wahlund, Henri, Lebreton, Miloch,
  Paulsson, Simon~Wedlund, Yang, Karlsson, \& {et al.}}]{eriksson_cold_2017}
Eriksson, A.~I., Engelhardt, I. A.~D., André, M., {et~al.} 2017, \aap

\bibitem[{Eriksson {et~al.}(2020)Eriksson, Gill, Johansson, \&
  Johansson}]{RosettaPSA}
Eriksson, A.~I., Gill, R., Johansson, E. P.~G., \& Johansson, F.~L. 2020, ESA
  Planetary Science Archive and Nasa Planetary Data System, {R}osetta RPC-LAP
  archive of derived plasma parameters from the ROSETTA mission

\bibitem[{Fahleson(1967)}]{fahleson_theory_1967}
Fahleson, U. 1967, Space Sci. Rev., 7, 238

\bibitem[{Galand {et~al.}(2016)Galand, Héritier, Odelstad, Henri, Broiles,
  Allen, Altwegg, Beth, Burch, Carr, Cupido, Eriksson, Glassmeier, Johansson,
  Lebreton, Mandt, Nilsson, Richter, Rubin, Sagnières, Schwartz, Sémon, Tzou,
  Vallières, Vigren, \& Wurz}]{galand_ionospheric_2016}
Galand, M., Héritier, K.~L., Odelstad, E., {et~al.} 2016, \mnras, 462, S331

\bibitem[{Garnier {et~al.}(2012)Garnier, Wahlund, Holmberg, Morooka, Grimald,
  Eriksson, Schippers, Gurnett, Krimigis, Krupp, Coates, Crary, \&
  Gustafsson}]{garnier_detection_2012}
Garnier, P., Wahlund, J.-E., Holmberg, M. K.~G., {et~al.} 2012, J. Geophys.
  Res., 117, 10202

\bibitem[{{Gilet} {et~al.}(2019){Gilet}, {Henri}, {Wattieaux}, {Myllys},
  {Randriamboarison}, {B{\'e}ghin}, \& {Rauch}}]{gilet_mutual_2019}
{Gilet}, N., {Henri}, P., {Wattieaux}, G., {et~al.} 2019, Frontiers in
  Astronomy and Space Sciences, 6, 16

\bibitem[{{Gilet, N.} {et~al.}(2020){Gilet, N.}, {Henri, P.}, {Wattieaux, G.},
  {Traor\'e, N.}, {Eriksson, A. I.}, {Valli\`eres, X.}, {Mor\'e, J.},
  {Randriamboarison, O.}, {Odelstad, E.}, {Johansson, F. L.}, \& {Rubin,
  M.}}]{Gilet2019cold}
{Gilet, N.}, {Henri, P.}, {Wattieaux, G.}, {et~al.} 2020, \aap, 640, A110

\bibitem[{Hansen {et~al.}(2016)Hansen, Altwegg, Berthelier, Bieler, Biver,
  Bockelée-Morvan, Calmonte, Capaccioni, Combi, de~Keyser, Fiethe, Fougere,
  Fuselier, Gasc, Gombosi, Huang, Le~Roy, Lee, Nilsson, Rubin, Shou, Snodgrass,
  Tenishev, Toth, Tzou, Wedlund, \& {Rosina Team}}]{hansen_evolution_2016}
Hansen, K.~C., Altwegg, K., Berthelier, J.-J., {et~al.} 2016, \mnras, 462, S491

\bibitem[{Henri {et~al.}(2017)Henri, Vallières, Hajra, Goetz, Richter,
  Glassmeier, Galand, Rubin, Eriksson, Nemeth, Vigren, Beth, Burch, Carr,
  Nilsson, Tsurutani, \& Wattieaux}]{henri_diamagnetic_2017}
Henri, P., Vallières, X., Hajra, R., {et~al.} 2017, \mnras, 469, S372

\bibitem[{Heritier {et~al.}(2017{\natexlab{a}})Heritier, Altwegg, Balsiger,
  Berthelier, Beth, Bieler, Biver, Calmonte, Combi, De~Keyser,
  {et~al.}}]{heritier_ion_2017}
Heritier, K.~L., Altwegg, K., Balsiger, H., {et~al.} 2017{\natexlab{a}},
  Monthly Notices of the Royal Astronomical Society, 469, S427

\bibitem[{Heritier {et~al.}(2018)Heritier, {Galand, M.}, {Henri, P.},
  {Johansson, F. L.}, {Beth, A.}, {Eriksson, A. I.}, {Valli\`eres, X.},
  {Altwegg, K.}, {Burch, J. L.}, {Carr, C.}, {Ducrot, E.}, {Hajra, R.}, \&
  {Rubin, M.}}]{heritier_plasma2018}
Heritier, K.~L., {Galand, M.}, {Henri, P.}, {et~al.} 2018, \aap, 618, A77

\bibitem[{Heritier {et~al.}(2017{\natexlab{b}})Heritier, Henri, Vallières,
  Galand, Odelstad, Eriksson, Johansson, Altwegg, Beth, Broiles, Burch, Carr,
  Cupido, Rubin, \& Vigren}]{heritier_vertical_2017}
Heritier, K.~L., Henri, P., Vallières, X., {et~al.} 2017{\natexlab{b}},
  \mnras, 469, S118

\bibitem[{Johansson {et~al.}(2020)Johansson, {Eriksson, A. I.}, {Gilet, N.},
  {Henri, P.}, {Wattieaux, G.}, {Taylor, M. G. G. T.}, {Imhof, C.}, \&
  {Cipriani, F.}}]{johansson_charging_2020}
Johansson, F.~L., {Eriksson, A. I.}, {Gilet, N.}, {et~al.} 2020, \aap, 642, A43

\bibitem[{Johansson {et~al.}(2017)Johansson, Odelstad, Paulsson, Harang,
  Eriksson, Mannel, Vigren, Edberg, Miloch, Simon~Wedlund, Thiemann, Eparvier,
  \& Andersson}]{johansson_rosetta_2017}
Johansson, F.~L., Odelstad, E., Paulsson, J. J.~P., {et~al.} 2017, Monthly
  Notices of the Royal Astronomical Society, 469, S626

\bibitem[{Karlsson {et~al.}(2017)Karlsson, Eriksson, Odelstad, Nilsson, Kullen,
  Lindqvist, Dickeli, Glassmeier, \& Richter}]{karlsson_rosetta_2017}
Karlsson, T., Eriksson, A.~I., Odelstad, E., {et~al.} 2017, Geophys. Res.
  Lett., 44

\bibitem[{Lindqvist {et~al.}(1994)Lindqvist, Marklund, \&
  Blomberg}]{lindqvist_plasma_1994}
Lindqvist, P.-A., Marklund, G.~T., \& Blomberg, L.~G. 1994, Space Sci. Rev.,
  70, 593

\bibitem[{Madanian {et~al.}(2016)Madanian, Cravens, Rahmati, Goldstein, Burch,
  Eriksson, Edberg, Henri, Mandt, Clark, Rubin, Broiles, \&
  Reedy}]{madanian_suprathermal_2016}
Madanian, H., Cravens, T.~E., Rahmati, A., {et~al.} 2016, Journal of
  Geophysical Research: Space Physics, 121

\bibitem[{Magnus \& Gudmundsson(2008)}]{magnus_digital_2008}
Magnus, F. \& Gudmundsson, J.~T. 2008, Review of Scientific Instruments, 79

\bibitem[{Morooka {et~al.}(2011)Morooka, Wahlund, Eriksson, Farrell, Gurnett,
  Kurth, Persoon, Shafiq, André, \& Holmberg}]{morooka_dusty_2011}
Morooka, M.~W., Wahlund, J.-E., Eriksson, A.~I., {et~al.} 2011, J. Geophys.
  Res., 116, A12221

\bibitem[{Mott-Smith \& Langmuir(1926)}]{mott-smith_theory_1926}
Mott-Smith, H.~M. \& Langmuir, I. 1926, Physical review, 28, 727

\bibitem[{{Myllys} {et~al.}(2019){Myllys}, {Henri, P.}, {Galand, M.},
  {Heritier, K. L.}, {Gilet, N.}, {Goldstein, R.}, {Eriksson, A. I.},
  {Johansson, F.}, \& {Deca, J.}}]{Myllys_plasma_prop2019}
{Myllys}, M., {Henri, P.}, {Galand, M.}, {et~al.} 2019, \aap, 630, A42

\bibitem[{Nilsson {et~al.}(2007)Nilsson, Lundin, Lundin, Barabash, Borg,
  Norberg, Fedorov, Sauvaud, Koskinen, Kallio, Riihelä, \&
  Burch}]{nilsson_rpc-ica:_2007}
Nilsson, H., Lundin, R., Lundin, K., {et~al.} 2007, Space Sci. Rev., 128, 671

\bibitem[{Nilsson {et~al.}(2017)Nilsson, {Stenberg Wieser}, Behar, Gunell,
  Wieser, Galand, Wdlund, Alho, Goetz, Yamauchi, Henri, Odelstad, \&
  Vigren}]{nilsson_evolution_2017}
Nilsson, H., {Stenberg Wieser}, G., Behar, E., {et~al.} 2017, \aap, 469, S252

\bibitem[{Nilsson {et~al.}(2020)Nilsson, Williamson, Bergman, Stenberg Wieser,
  Wieser, Behar, Eriksson, Johansson, Richter, \&
  Goetz}]{nilsson_average2020mnras}
Nilsson, H., Williamson, H., Bergman, S., {et~al.} 2020, \mnras, 498, 5263

\bibitem[{Odelstad {et~al.}(2015)Odelstad, Eriksson, Edberg, Johansson, Vigren,
  André, Tzou, Carr, \& Cupido}]{odelstad_evolution_2015}
Odelstad, E., Eriksson, A.~I., Edberg, N. J.~T., {et~al.} 2015, Geophys. Res.
  Lett., 42, 10,126

\bibitem[{{Odelstad} {et~al.}(2018){Odelstad}, {Eriksson}, {Johansson},
  {Vigren}, {Henri}, {Gilet}, {Heritier}, {Valli{\`e}res}, {Rubin}, \&
  {Andr{\'e}}}]{odelstad_ion_2018}
{Odelstad}, E., {Eriksson}, A.~I., {Johansson}, F.~L., {et~al.} 2018, Journal
  of Geophysical Research (Space Physics), 123, 5870

\bibitem[{Odelstad {et~al.}(2017)Odelstad, Stenberg-Wieser, Wieser, Eriksson,
  Nilsson, \& Johansson}]{odelstad_measurements_2017}
Odelstad, E., Stenberg-Wieser, G., Wieser, M., {et~al.} 2017, \mnras, 469, S568

\bibitem[{Ruhunusiri {et~al.}(2020)Ruhunusiri, Howes, \&
  Halekas}]{ruhunusiri_turbulence_67P2020JGR}
Ruhunusiri, S., Howes, G.~G., \& Halekas, J.~S. 2020, Journal of Geophysical
  Research: Space Physics, 125, e2020JA028100, e2020JA028100
  10.1029/2020JA028100

\bibitem[{Sagalyn {et~al.}(1963)Sagalyn, Smiddy, \&
  Wisnia}]{sagalyn_measurement_1963}
Sagalyn, R.~C., Smiddy, M., \& Wisnia, J. 1963, J. Geophys. Res., 68, 199

\bibitem[{Taylor {et~al.}(2017)Taylor, Altobelli, Buratti, \&
  Choukroun}]{taylor_rosetta_2017}
Taylor, M. G. G.~T., Altobelli, N., Buratti, B.~J., \& Choukroun, M. 2017,
  Philosophical Transactions of the Royal Society of London A: Mathematical,
  Physical and Engineering Sciences, 375

\bibitem[{Trotignon {et~al.}(2007)Trotignon, Michau, Lagoutte, Chabassière,
  Chalumeau, Colin, DÈcrÈau, Geiswiller, Gille, Grard, Hachemi, Hamelin,
  Eriksson, Laakso, Lebreton, Mazelle, Randriamboarison, Schmidt, Smit,
  Telljohann, \& Zamora}]{trotignon_rpc-mip:_2007}
Trotignon, J.-G., Michau, J.~L., Lagoutte, D., {et~al.} 2007, Space Sci. Rev.,
  128, 713

\bibitem[{Vigren {et~al.}(2016)Vigren, Altwegg, Edberg, Eriksson, Galand,
  Henri, Johansson, {E. Odelstad}, Tzou, \&
  Valliéres}]{vigren_model-observation_2016}
Vigren, E., Altwegg, K., Edberg, N. J.~T., {et~al.} 2016, The Astronomical
  Journal, 152, 59

\bibitem[{Vigren {et~al.}(2017)Vigren, André, Edberg, Engelhardt, Eriksson,
  Galand, Goetz, Henri, Heritier, Johansson, Nilsson, Odelstad, Rubin,
  {Stenberg Wieser}, Tzou, \& Vallières}]{vigren_effective_2017}
Vigren, E., André, M., Edberg, N. J.~T., {et~al.} 2017, \mnras, 469, S142

\bibitem[{Vigren \& Eriksson(2017)}]{vigren_1d_2017}
Vigren, E. \& Eriksson, A.~I. 2017, The Astronomical Journal, 153, 150

\bibitem[{{Wattieaux} {et~al.}(2019){Wattieaux}, {Gilet}, {Henri},
  {Valli{\`e}res}, \& {Bucciantini}}]{wattieaux_rpcmip_model_2019}
{Wattieaux}, G., {Gilet}, N., {Henri}, P., {Valli{\`e}res}, X., \&
  {Bucciantini}, L. 2019, \aap, 630, A41

\bibitem[{Wattieaux {et~al.}(2020)Wattieaux, Henri, Gilet, Vallieres, \&
  Deca}]{wattieaux_plasma_2020}
Wattieaux, G., Henri, P., Gilet, N., Vallieres, X., \& Deca, J. 2020, \aap

\end{thebibliography}

\end{document}